\documentclass[aps,prl,twocolumn,showpacs,superscriptaddress,groupedaddress]{revtex4}  
\usepackage{graphicx}  
\usepackage{dcolumn}   
\usepackage{bm}        
\usepackage{amssymb}   
\usepackage{amsmath}
\usepackage{float}
\usepackage{slashed}
\usepackage{hyperref}
\usepackage{color} 

\newcommand{\code}[1]{\texttt{#1}}

\begin{document}

\title{Self-interacting hidden sector dark matter, small scale galaxy  structure anomalies, and a dark force}

\author{Amin Aboubrahim}
\thanks{{\scriptsize Email}: \href{mailto:abouibrahim.a@northeastern.edu}{abouibrahim.a@northeastern.edu}; {\scriptsize ORCID}: \href{https://orcid.org/0000-0002-1110-4265}{0000-0002-1110-4265}}
\affiliation{Department of Physics, Northeastern University, Boston, MA 02115-5000, USA}
\author{Wan-Zhe Feng}
\thanks{{\scriptsize Email}: \href{mailto:vicf@tju.edu.cn}{vicf@tju.edu.cn}; {\scriptsize ORCID}: \href{https://orcid.org/0000-0003-2488-0041}{0000-0003-2488-0041}}
\affiliation{Center for Joint Quantum Studies and Department of Physics, School of Science, Tianjin University, Tianjin 300350, PR. China}
\author{Pran Nath}
\thanks{{\scriptsize Email}: \href{mailto:p.nath@northeastern.edu}{p.nath@northeastern.edu}; {\scriptsize ORCID}: \href{https://orcid.org/0000-0001-9879-9751}{0000-0001-9879-9751}}
\affiliation{Department of Physics, Northeastern University, Boston, MA 02115-5000, USA}
\author{Zhu-Yao Wang}
\thanks{{\scriptsize Email}: \href{mailto:wang.zhu@northeastern.edu}{wang.zhu@northeastern.edu}; {\scriptsize ORCID}: \href{https://orcid.org/0000-0001-5398-7302}{0000-0001-5398-7302}}
\affiliation{Department of Physics, Northeastern University, Boston, MA 02115-5000, USA}

\date{\today}

\begin{abstract}
The short distance behavior of dark matter (DM) at galaxy scales exhibits several features not explained by the typical cold dark matter (CDM) with velocity-independent cross-section. We discuss a particle physics model with a hidden sector interacting feebly with the visible sector where a dark fermion self-interacts via a dark force with a light dark photon as the mediator. We study coupled Boltzmann equations involving two temperatures, one for each sector. 
We fit the velocity-dependent DM cross-section to the data from scales of dwarf galaxies to clusters consistent with relic density constraint. 
\end{abstract}

\maketitle

\section{Introduction} 
While the $\Lambda$CDM model works very well at large scales, several issues have arisen recently concerning weakly interacting massive particles (WIMPs) as CDM with regards to physics at galaxy scales.
Some of  these are described as the cusp-core, the missing satellites,
and  the too-big-to-fail (TBTF) anomalies.
 A comprehensive review of these issues can be found in the paper by Tulin and Yu~\cite{Tulin:2017ara}.
There are various suggestions on how to overcome some of these anomalies such as 
 using complex dynamics and baryonic physics along 
with WIMP simulations~\cite{Governato:2012fa}, ultralight axions~\cite{Kim:2015yna,Hui:2016ltb,Halverson:2017deq} as alternative to WIMPs 
and self-interacting dark matter (SIDM).  The last suggestion first made by Spergel and Steinhardt~\cite{Spergel:1999mh} has recently attracted considerable interest~\cite{Vogelsberger:2012ku,Rocha:2012jg,Peter:2012jh,Zavala:2012us,Elbert:2014bma,Vogelsberger:2014pda,Fry:2015rta,Dooley:2016ajo,Buckley:2009in,Loeb:2010gj,Tulin:2012wi,Tulin:2013teo,Schutz:2014nka,Bringmann:2016din}.
The SIDM models allow for a fit to the data from the scales of dwarf galaxies, where the SIDM acts like
a collisional fluid, to  galaxy clusters, where SIDM becomes collisionless. Data from dwarf galaxy scales to galaxy clusters~\cite{Robertson:2018anx,Postman, Sagunski:2020spe, Andrade:2020lqq,Elbert:2016dbb} will be collectively denoted as DGC in this work.
   Most of the analyses to fit the DGC data use Yukawa interactions to model self-interactions, where the computation of the dark matter relic density presents a challenge~\cite{Kaplinghat:2015aga}. 
 
 There is currently a significant amount of data from dwarf galaxies to galaxy clusters 
 and it is of interest to see if such data  hides any clues to the nature of dark matter 
 which may allow us to discriminate among various DM models.  One important  indicator here is the possible velocity dependence of dark matter  cross sections. It turns out that the velocity dependence is a possible way to differentiate  a class of SIDM models from CDM. Thus the SIDM models based on particle
  exchange produce a scattering cross section which goes like $1/v^4$ as in Rutherford 
  scattering which gives a negligible cross section for large velocities and SIDM in this 
  region acts like a collisionless fluid.   This is the situation for galaxy clusters 
  where  $v$ tends to be as large as $1000$ km/s or larger and $\sigma/m$ has an upper limit which
 is estimated to be maximally 1 cm$^{2}$/g~\cite{Tulin:2017ara,Kaplinghat:2015aga,Robertson:2018anx,Postman}  and
as low as 0.1 cm$^{2}$/g~\cite{Elbert:2016dbb,Sagunski:2020spe} to   0.065 cm$^{2}$/g at $95\%$ CL~\cite{Andrade:2020lqq}.  Here one may fit the data either by CDM or by SIDM. 
   However,  for mid-size galaxies such as the Milky Way and low surface brightness galaxies
  where $\langle v\rangle$ lies in the range
   $\sim$ 80$-$200 km/s, fit to data indicates $\sigma/m\sim$ 0.5$-$5 cm$^2$/g 
   and for dwarf galaxies where $\langle v\rangle \sim$ 10$-$100 km/s,  $\sigma/m$   lies in the 
   range 1$-$50 cm$^2$/g~\cite{Tulin:2017ara,Kaplinghat:2015aga}.  
   Thus one finds that for velocities smaller than those in the galaxy cluster range, CDM and SIDM behave differently since in this region SIDM becomes a collisional fluid and
 helps resolve  the cusp-core and the TBTF anomalies.
 Velocity dependence of SIDM is the underlying reason for the transition of SIDM from one form
 to the other,   and 
 the desired velocity dependence appears naturally in 
    SIDM models where the dark matter is composed of dark fermions of mass $m_D$ and the self-interaction
    arises from a dark force due to the exchange of a dark photon of mass $m_{\gamma'}$, where $m_{\gamma'}/m_D\ll 1$.
    
 It is of interest to construct particle physics models which can explain the DGC data along with satisfying the relic  density constraint. Since the hidden sector and the visible sector in general will have different temperatures~\cite{Feng:2008mu,Chu:2011be,Ackerman:mha,Foot:2014uba,Foot:2016wvj,Hambye:2019dwd}, a proper 
 analysis of the coupled hidden and visible sectors requires study of Boltzmann equations involving temperatures of both the hidden and the visible sectors, which we carry out in the analysis below. 
 
\section{Hidden sector dark matter, a dark force and a feeble coupling to the visible sector}
In this work we construct  models where the dark matter particles have feeble interactions with the visible sector 
  and  are produced in the early universe by the freeze-in mechanism~\cite{Hall:2009bx,Aboubrahim:2019kpb,Aboubrahim:2020wah,Koren:2019iuv,Du:2020avz}.
  Specifically we  consider an extended standard model with
 a hidden sector which has  matter and gauge fields with a $U(1)_X$ gauge invariance which has mixings 
 with the visible sector $U(1)_Y$  via gauge kinetic~\cite{Holdom:1985ag,Holdom:1991,Dutra:2018gmv} and Stueckelberg mass mixings~\cite{Kors:Nath,st-mass-mixing,Feldman:2007wj, WZFPN,Aboubrahim:2019qpc}.
 The relevant part of the Lagrangian of the extended  model is 
 \begin{align}
 \mathcal{L}=& -\frac{1}{4} C^{\mu\nu} C_{\mu\nu} - g_X \bar D \gamma^\mu DC_\mu  +m_D \bar D D \nonumber\\
 &- \frac{\delta}{2} C^{\mu\nu} B_{\mu\nu} - \frac{1}{2}(M_1C_\mu + M_2 B_\mu + \partial_\mu \sigma)^2,
 \end{align}
 where $C_\mu$ is the gauge field of $U(1)_X$,  $B_\mu$ is the gauge field for the $U(1)_Y$, $\sigma$ is an 
 axion field which gives mass to $C_\mu$ and is absorbed in the unitary gauge,  
 $D$ is a Dirac fermion which is charged under $U(1)_X$, $\delta$ is the kinetic
 mixing parameter,  $M_1$ and $M_2$  are the mass parameters in the Stueckelberg mass mixing. The diagonalization of the gauge boson
 mass matrix along with the mass matrix arising from the spontaneous breaking of the Higgs boson in $SU(2)\times U(1)_Y$ gives the following mass eigenstates:  the photon $(\gamma)$, the $Z$ boson, and $Z'(\gamma')$. Because the mass
 of the third neutral boson would turn out to be in MeV region we will refer to it as a dark photon or $\gamma'$ which, however,
 is unstable and decays.
 
\section{Deduction of self-Consistent Two-Temperature Boltzmann  Equations} 
We give in this section a deduction of the temperature-dependent coupled Boltzmann equations of $D$ and $\gamma'$ and the evolution equation of $\eta=T/T_h$, with $T\,(T_h)$ being the visible (hidden) sector temperature. One then obtains the set of basic equations that govern the evolution of the particle number densities in the visible and hidden sectors when the two sectors have different bath temperatures. These equations, solved simultaneously, are essential for a proper analysis of the coupled  visible sector-hidden sector system in such a situation.  One consequence of coupling of the 
 visible and the hidden sectors is that the  entropies in the hidden and the visible sectors
 are not individually preserved but it is only their sum which is a constraint imposed in the
 analysis. In this work, we use the hidden sector temperature  as the clock and the temperature in the 
 visible sector is related to the hidden sector via the function $\eta$. 
 
 We begin by considering the two Friedman equations for a flat universe
 \begin{align} 
 \label{FR1}
 H^2 &= \frac{8\pi G_N} {3} \rho, \\
 \frac{ \ddot{a}}{a}&= - \frac{4\pi G_N}{3} (\rho+3 p),
 \label{FR2}
 \end{align}
 where $G_N$ is Newton's gravitational constant, $\rho$ and $p$ are the energy density and pressure, respectively. 
 Differentiating Eq.~(\ref{FR1}) and using  Eqs.~(\ref{FR1}) and (\ref{FR2}), we can deduce the result 
  \begin{align} 
  \frac{d\rho}{dt}+ 3H(\rho + p) =0.
  \label{FR3}
 \end{align}
 As noted above we will use $T_h$ as the clock and we can then obtain from Eq.~(\ref{FR3}) 
 the following relation
  \begin{align} 
 \frac{dT_h}{dt}= - \frac{4\zeta\rho}{\frac{d\rho}{dT_h}} H, 
 \label{FR4}
 \end{align}
 where $\zeta=\frac{3}{4} (1+ p/\rho)$. Here $\zeta=1$ is for the radiation dominated era and 
 $\zeta=3/4$ for the matter dominated universe. 
 We wish to determine $d\rho_v/dT_h$ in terms of $d\rho_h/dT_h$ (the subscripts $v$ and $h$ correspond to the visible and hidden sectors, respectively).
 We begin by considering the equation obeyed by $\rho_h$:
 \begin{align}
 \frac{d\rho_h}{dt} + 3 H(\rho_h+ p_h) =j_h,
 \label{a1}
 \end{align}
 where $\rho_h$ is the energy density, $p_h$ is the pressure in the hidden sector and $j_h$ is the source
 term in the hidden sector and arises from freeze-in.
 Next, we write 
  \begin{align}
  \frac{d\rho_h}{dt}=   
 \frac{dT_h}{dt} \frac{d\rho_h}{dT_h},
  \label{a3} 
 \end{align}
and upon using Eqs.~(\ref{FR4}), (\ref{a1}) and (\ref{a3}), 
we get
 \begin{align}
\rho \frac{d\rho_h}{dT_h} =\left(\frac{\zeta_h}{\zeta} \rho_h-\frac{j_h}{4H\zeta}\right) \frac{d\rho}{dT_h},
 \label{a7} 
\end{align}
where $\zeta_h= \frac{3}{4} (1+p_h/\rho_h)$ and  interpolates between $\zeta_h=1$ for radiation dominance and $\zeta_h=3/4$ for matter dominance in the hidden sector. We  note that  
  since $\rho= \rho_v + \rho_h$,  we have $d\rho/dT_h=d\rho_v/dT_h+d\rho_h/dT_h$ and together with
Eq.~(\ref{a7}), we can 
 solve for $d\rho_v/dT_h$ in terms of $d\rho_h/dT_h$   and get 
\begin{align}
 \frac{d\rho_v}{dT_h}   =   \frac{\zeta \rho_v+ \rho_h( \zeta-\zeta_h)+ j_h/(4H)}{\zeta_h\rho_h- j_h/(4H)}     \frac{d\rho_h}{dT_h}.
 \label{a10} 
\end{align}
 Next, we use Eq.~(\ref{a10}) to obtain an equation for $d\eta/dT_h$.  
Now $\eta(T_h)$ enters only in $d\rho_v/dT_h$  and to
 compute it we use 
\begin{align}
 \frac{dT}{dT_h}&= 
 \eta + T_h\eta', 
 \label{a14}
 \end{align}
where $\eta'= d\eta/dT_h$.

In the analysis, we will use the constraint that the total entropy $S=s R^3$ is conserved which gives $ds/dt + 3 Hs=0$. 
Here  $s=s_v+ s_h$, where $s_v$  depends on $T$ and $s_h$  on $T_h$ so that 
\begin{align}
s&=\frac{2\pi^2}{45}\left(h_{\rm eff}^h T_h^3+h_{\rm eff}^v T^3\right),
\label{entropy}
\end{align}
where $h^v_{\rm eff}\,(h^h_{\rm eff})$ is the visible (hidden) effective entropy degrees of freedom.
 The Hubble parameter also depends on both $T$ and $T_h$ as can be seen
from the Friedman equation 
\begin{align}
H^2= \frac{8\pi G_N}{3} (\rho_v(T)  +\rho_h(T_h)),
\label{hubble}
\end{align}
where $\rho_v(T) (\rho_h(T_h))$ is the energy density in the visible (hidden) sector at temperature $T(T_h)$ and given by
\begin{align}
\rho_v&=\frac{\pi^2}{30}g_{\rm eff}^v T^4, ~~
\rho_h=\frac{\pi^2}{30}g_{\rm eff}^h T_h^4.
\label{rho-1} 
\end{align}
 $g^v_{\rm eff}, h^v_{\rm eff}$   are functions of $T$ and we use the fits given in~\cite{Kolb:1990vq,Gondolo:1990dk,Gelmini:1990je} to parametrize them while 
 $g^h_{\rm eff}, h^h_{\rm eff}$ are functions of $T_h$ and we use temperature dependent 
integrals given in~\cite{Hindmarsh:2005ix} to parametrize them.  

Using Eq.~(\ref{rho-1}) and Eq.~(\ref{a14}) we get 
 \begin{align}
  \label{a16} 
\frac{d\rho_v}{dT_h}&=
 A_v+ B_v \eta',
 \end{align}
 where $A_v$ and $B_v$ are given by
 \begin{align}
\label{y4}
A_v&=\frac{\pi^2}{30}\left(\frac{dg_{\rm eff}^v}{dT}\eta^5 T_h^4+4g_{\rm eff}^v\eta^4 T_h^3\right),\\
B_v&=\frac{\pi^2}{30}\left(\frac{dg_{\rm eff}^v}{dT}\eta^4 T_h^5+4g_{\rm eff}^v\eta^3 T_h^4\right). 
\label{y5}
\end{align}
Using Eqs.~(\ref{a10}) and (\ref{a16}), we get 
\begin{align}
   A_v + B_v \eta'=  \frac{\zeta \rho_v+ \rho_h( \zeta-\zeta_h)+ j_h/(4H)}{\zeta_h\rho_h- j_h/(4H)}     \frac{d\rho_h}{dT_h},
  \label{a18}    
 \end{align}
which is solved for $\eta'$ to get Eq.~(\ref{y3}). 

Now $\rho_h$ and $p_h$, which enter in the definition of $\zeta_h$, are determined
in terms of $\rho_{\gamma'}, p_{\gamma'}, \rho_D, p_D$ so that $\rho_h= \rho_{\gamma'} + \rho_D$ and $p_h= p_{\gamma'} + p_D$, 
where 
$\rho_{\gamma'}$ and $p_{\gamma'}$ are given by
\begin{align}
\begin{aligned}
\rho_{\gamma'}&=\frac{g_{\gamma'} T^4}{2\pi^2} 
\int_{x_{\gamma'}}^{\infty} \frac{x^3 dx}{ e^{x} -1}, \\
p_{\gamma'}&=\frac{g_{\gamma'} T^4}{6\pi^2} 
\int_{x_{\gamma'}}^{\infty} \frac{(x^2-x^2_{\gamma'})dx}{ e^{x} -1}.
\label{integral-a}
\end{aligned}
\end{align}
Similarly, for the $\rho_D$ and $p_D$, we have 
\begin{align}
\begin{aligned}
\rho_{D}&=\frac{g_{D} T^4}{2\pi^2} 
\int_{x_{D}}^{\infty} \frac{x^3 dx}{ e^{x} +1}, \\
p_{D}&=\frac{g_{D} T^4}{6\pi^2} 
\int_{x_{D}}^{\infty} \frac{(x^2-x^2_{\gamma'})dx}{ e^{x} +1}.
\label{Integral-b} 
\end{aligned}
\end{align}
Here $g_{\gamma'}=3$ and $g_D=4$ and we have used the natural unit system $c=k_B=1$, with
$x_{\gamma'}= m_{\gamma'}/T_h$ and $x_D= m_D/T_h$.
In the computation of $\zeta$ one needs  $\rho=\rho_v+ \rho_h$ and $p=p_v+p_h$
where the computation of $\rho_v$ and $p_v$ is done numerically as discussed in the text.

Next, we  discuss the Boltzmann equations for the number densities of the dark fermions $D$ and 
of the dark photons $\gamma'$ using the hidden sector temperature $T_h$ as the clock. 
In this case, for the $D$ fermions, we have 
\begin{align}
\label{n1}
\frac{dn_D}{dt} + 3 Hn_D=  
&\Big[\langle\sigma v\rangle_{D\bar{D}\to i\bar{i}}(T)n_D^{\rm eq}(T)^2 \nonumber \\
&-\langle\sigma v\rangle_{D\bar{D}\to\gamma'\gamma'}(T_h)n_D(T_h)^2 \nonumber \\
&+\langle\sigma v\rangle_{\gamma'\gamma'\to D\bar{D}}(T_h)n_{\gamma'}(T_h)^2\Big].
\end{align}

In a similar fashion the Boltzmann equation for $n_{\gamma'}$ is given by 
\begin{align}
\label{n2}
\frac{dn_{\gamma'}}{dt} + 3 Hn_{\gamma'}= 
&\Big[\langle\sigma v\rangle_{D\bar{D}\to\gamma'\gamma'}(T_h)n_D(T_h)^2 \nonumber \\
&-\langle\sigma v\rangle_{\gamma'\gamma'\to D\bar{D}}(T_h)n_{\gamma'}(T_h)^2 \nonumber \\
&+\langle\sigma v\rangle_{i\bar{i}\to\gamma'}(T)n_i^{\rm eq}(T)^2 \nonumber \\
&-
{\langle\Gamma_{\gamma'\to i\bar{i}}(T_h)\rangle }n_{\gamma'}(T_h)\Big].
\end{align} 
In Eqs.~(\ref{n1}) and~(\ref{n2}), the thermally averaged cross-section and decay widths are given by
\begin{widetext}
\begin{equation}
\langle\sigma v\rangle^{a\bar{a}\to bc}(T)=\frac{1}{8 m^4_a T K^2_2(m_a/T)}\int_{4m_a^2}^{\infty} ds ~\sigma(s) \sqrt{s}\, (s-4m_a^2)K_1(\sqrt{s}/T),
\end{equation}
\end{widetext}
and
\begin{equation}
\langle\Gamma_{X\to i\bar{i}}(T)\rangle=\Gamma_{X\to i\bar{i}}\frac{K_1(m_{X}/T)}{K_2(m_{X}/T)},
\end{equation}
with $K_1$ and $K_2$ being the modified Bessel function of the second kind and degrees one and two, respectively. Note that standard thermal averaging is used in the dark sector since the $D$ fermions enter immediately in self-equilibration (see discussion in the next section). Deviations from this scenario may occur and must be treated with care~\cite{Binder:2017rgn,DEramo:2020gpr}.

In Eq.~(\ref{n1}) and Eq.~(\ref{n2}),  we will use $T_h$ as the reference temperature and replace $t$ by $T_h$. We then analyze the evolution of 
 $n_D$, $n_{\gamma'}$ and $\eta$ as a function of $T_h$. For the computation of the relic density, 
 it is more convenient to deal directly with particle yields
  defined by $Y_a = n_a/s$ for a particle species $a$ with number density $n_a$. 
  We assume that the dark particles $D, \gamma'$ are feeble and there is no initial abundance and that they are initially
  produced only via freeze-in processes such as $i\bar{i}\to D\bar D$, $i\bar{i}\to \gamma'$, where $i$ refers to standard model
  particles. However, $D$ and $\gamma'$ have interactions such as $D\bar D\to \gamma' \gamma'$ 
  within the hidden sector which, in our case, are not feeble.  
  The Boltzmann equations  
  for the yields $Y_D$ and $Y_{\gamma'}$ and the evolution $\eta$ then take the form
\begin{align}
\label{y1}
\frac{dY_D}{dT_h}=&-\frac{s}{H}\Big(\frac{d\rho_h/dT_h}{4\zeta_h\rho_h-j_h/H}\Big)\Big[\langle\sigma v\rangle_{D\bar{D}\to i\bar{i}}(T)Y_D^{\rm eq}(T)^2\nonumber\\
&-\langle\sigma v\rangle_{D\bar{D}\to\gamma'\gamma'}(T_h)Y^2_D+\langle\sigma v\rangle_{\gamma'\gamma'\to D\bar{D}}(T_h)Y^2_{\gamma'}\Big],\\
\label{y2}
\frac{dY_{\gamma'}}{dT_h}=&-\frac{s}{H}\left(\frac{d\rho_h/dT_h}{4\zeta_h\rho_h-j_h/H}\right)\Big[\langle\sigma v\rangle_{D\bar{D}\to\gamma'\gamma'}(T_h)Y^2_D\nonumber\\
&-\langle\sigma v\rangle_{\gamma'\gamma'\to D\bar{D}}(T_h)Y^2_{\gamma'}-\frac{1}{s}\langle\Gamma_{\gamma'\to i\bar{i}}(T_h)\rangle Y_{\gamma'}\nonumber \\
&+\langle\sigma v\rangle_{i\bar{i}\to\gamma'}(T)Y_i^{\rm eq}(T)^2\Big],\\
\frac{d\eta}{dT_h}= &- \frac{A_v}{B_v} +  \frac{\zeta \rho_v+ \rho_h( \zeta-\zeta_h)+ j_h/(4H)}{\zeta_h\rho_h- j_h/(4H)}     \frac{\frac{d\rho_h}{dT_h}}{B_v},
\label{y3}
\end{align}
where 
\begin{align}
\label{y6}
j_h=&\sum_i \Big[2Y^{\rm eq}_i(T)^2 J(i\bar{i}\to D\bar{D})(T)\nonumber\\
&+Y^{\rm eq}_i(T)^2 J(i\bar{i}\to \gamma')(T)\Big]s^2\nonumber\\
&-Y_{\gamma'}J(\gamma'\to e^+ e^-)(T_h)s, \\
\label{y7}
Y^{\rm eq}_i=&\frac{n_i^{\rm eq}}{s}=\frac{g_i}{2\pi^2 s}m_i^2 T K_2(m_i/T).
\end{align}
Here $g_i$ is the number of degrees of freedom of particle $i$ and mass $m_i$ and the source functions $J$ are discussed in the Appendix. Note that in Eq.~(\ref{y2}) there are contributions one can add on the right hand side which involve processes $i\bar{i}\to \gamma' \gamma,
\gamma'Z, \gamma'\gamma'$. However, their contributions are relatively small compared to $i~\bar{i}\to \gamma'$. 
 
 The entropy density and the Hubble parameter given by Eqs.~(\ref{entropy}) and~(\ref{hubble}) can be rewritten as
 \begin{align}
 s&=
\frac{2\pi^2}{45}  h_{\rm eff}  T_h^3,~~~\text{and}~~~  
 H^2=
   \frac{8\pi G_N}{3}  \frac{\pi^2}{30} g_{\rm eff} T_h^4,   
        \end{align}
where the total entropy and energy density effective degrees of freedom are defined as       
        \begin{align}
    h_{\rm eff}&= h^h_{\rm eff}  + \eta^3 h^v_{\rm eff},~~~\text{and}~~~g_{\rm eff}= g^h_{\rm eff}  + \eta^4 g^v_{\rm eff}.   
        \end{align}        
Thus, the ratio $s/H$ that appears in Eqs.~(\ref{y1}) and (\ref{y2}) can be written as
 \begin{align}
 \frac{s}{H}&= 
   \frac{ 2 \sqrt 2 \pi}{  \sqrt{45}}    \frac{h_{\rm eff}}{ \sqrt{g_{\rm eff}}} M_{\rm Pl}  T_h,  
  \label{sH}
  \end{align}
 where
 $M_{\rm Pl}\equiv \sqrt{ \frac{1}{ 8\pi G_N}}  = 2.4 \times 10^{18}$ GeV.

In the dark sector, the effective degrees of freedom include those for the  dark photon and for the dark fermion so that
\begin{align}
g^h_{\rm eff}&= g^{\gamma'}_{\rm eff} +\frac{7}{8}  g^D_{\rm eff},~~~\text{and}~~~h^h_{\rm eff}= h^{\gamma'}_{\rm eff} + \frac{7}{8} h^D_{\rm eff}.
\end{align}

At  temperature $T_h$, $g_{\rm eff}$ and  $h_{\rm eff}$  for the particles $\gamma'$ and $D$ are 
given by 
\begin{widetext}
\begin{equation}
\begin{aligned}
g^{\gamma'}_{\rm eff}& = \frac{45}{\pi^4} \int_{x_{\gamma'}}^{\infty} \frac{\sqrt{x^2-x_{\gamma'}^2} }{e^x-1 } x^2 dx,~~~\text{and}~~~h^{\gamma'}_{\rm eff}= \frac{45}{4\pi^4} \int_{x_{\gamma'}}^{\infty} \frac{\sqrt{x^2-x_{\gamma'}^2} }{e^x-1 } 
(4x^2-x_{\gamma'}^2) dx, \\
g^{D}_{\rm eff}& = \frac{60}{\pi^4} \int_{x_{D}}^{\infty} \frac{\sqrt{x^2-x_{D}^2} }{e^x+1 } x^2 dx,~~~\text{and}~~~h^{D}_{\rm eff}= \frac{15}{\pi^4} \int_{x_{D}}^{\infty} \frac{\sqrt{x^2-x_{D}^2} }{e^x+1 } 
(4x^2-x_D^2) dx,
\end{aligned}
\label{hdof}
\end{equation}
\end{widetext}
where $x_{\gamma'}$ and $x_D$ are as defined after Eq.~(\ref{Integral-b}).
 We note that in the limit 
$x_{\gamma'}\to 0$  one has $g^{\gamma'}_{\rm eff}= h^{\gamma'}_{\rm eff}\to 3$
  and when $x_D\to 0$ one has $g^{D}_{\rm eff}= h^{D}_{\rm eff}\to 4$.

\begin{table}
\caption{\label{tab1} 
The benchmarks used in the analysis
where we set $M_2=0$ and $\delta$ is in units of $10^{-9}$.}
\begin{ruledtabular}
\begin{tabular}{cccccc}
&Model & $m_D$ (GeV) & $M_1$ (MeV) & $g_X$ & $\delta$\\
\hline
&(a) & 1.50 & 1.20 & 0.016 & $28$ \\
&(b) & 2.0 & 1.22 & 0.014 & $4.0$\\
&(c) & 2.16 & 1.13 & 0.015 & $4.7$\\
&(d) & 3.2 & 1.77 & 0.018 & $3.8$ \\
&(e) & 3.26 & 1.99 & 0.018 & $3.5$\\
&(f) & 4.0 & 2.20 & 0.020 & $3.6$ \\
\hline
\hline
&Model  & $\sigma/m_D$ (cm$^2$/g) & $\Omega h^2$ & $\Gamma_{\gamma'\to e^+ e^-}$ (GeV) & $\tau$ (ms) \\
\hline
& (a) & 2.48 & 0.1215 & $1.4\times 10^{-21}$ & 0.49 \\
& (b) & 1.97 & 0.1233 & $2.9\times 10^{-23}$ & 22.7 \\
& (c) & 3.69 & 0.1218 & $3.0\times 10^{-23}$ & 21.8 \\
& (d) & 1.79 & 0.1191 & $4.9\times 10^{-23}$ & 13.4  \\
& (e) & 1.24 & 0.1185 & $4.8\times 10^{-23}$ & 13.8 \\
& (f) & 1.43 & 0.1229 & $5.6\times 10^{-23}$ & 11.7 \\
\end{tabular}
\end{ruledtabular}
\end{table}

\section{Dark freeze-out, relic density, and fits to DGC data}
We give now a numerical analysis based on the formalism of the preceding section.
 In Table~\ref{tab1} we give a set of six benchmarks which satisfy
 the relic density constraint and where the dark photon decays before the Big Bang Nucleosynthesis (BBN). The values of $\sigma/m$ at low velocities for these model points lie in the range (1.2$-$3.7) cm$^2$/g which are needed to explain the short distance structure  of dark matter at galaxy scales.  The calculation of the relic density requires solving the set of stiff differential equations, Eqs.~(\ref{y1})$-$(\ref{y3}), and integrating the yield of $D$ fermions to present day temperature to obtain $Y_D^0$. In solving the coupled system, the effective number of degrees of freedom for the hidden sector, $g_{\rm eff}^h$ and $h_{\rm eff}^h$, are determined from the set of equations, Eq.~(\ref{hdof}), while those for the visible sector, $g_{\rm eff}^v$ and $h_{\rm eff}^v$, are read from tabulated results in \code{micrOMEGAs} obtained from Refs.~\cite{Gondolo:1990dk,Gelmini:1990je}. The relic density of $D$ is then determined by using
\begin{align}
\Omega h^2 = \frac{m_D Y^0_D s_0 h^2}{\rho_c}, 
\label{relic}
\end{align}
where $\rho_c$ is the critical density, $s_0$ is today's entropy density and $h=0.678$. 
 
\begin{figure}
   \includegraphics[width=0.49\textwidth]{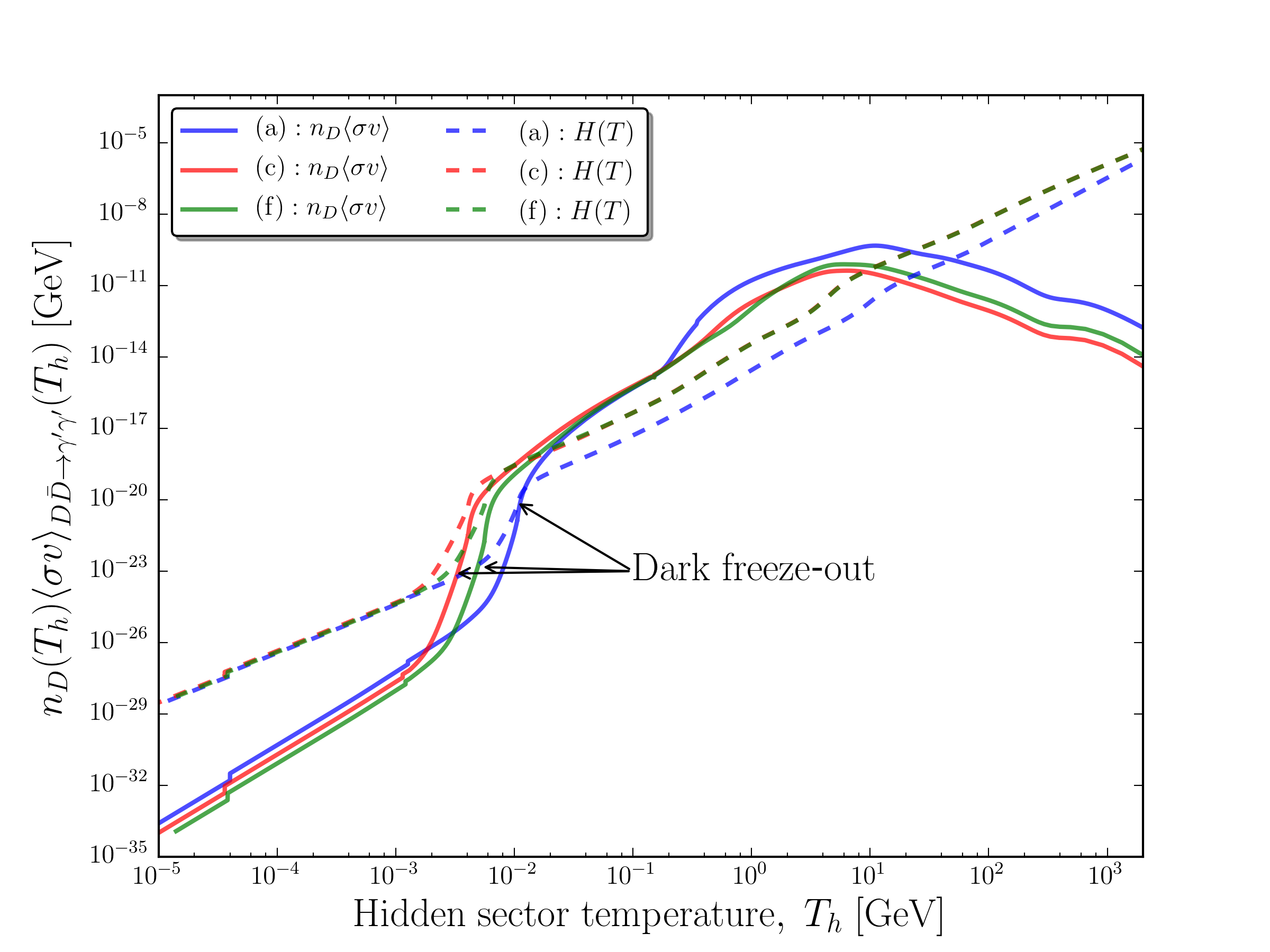} 
    \caption{A display of dark freeze-out showing a plot of 
    $n_D\langle\sigma v\rangle_{D\bar{D}\to\gamma'\gamma'}$ (solid line) and $H(T)$ (dashed line) versus $T_h$ for three benchmarks of Table~\ref{tab1}.}
\label{fig1}
\end{figure}

\begin{figure}
 \includegraphics[width=0.5\textwidth]{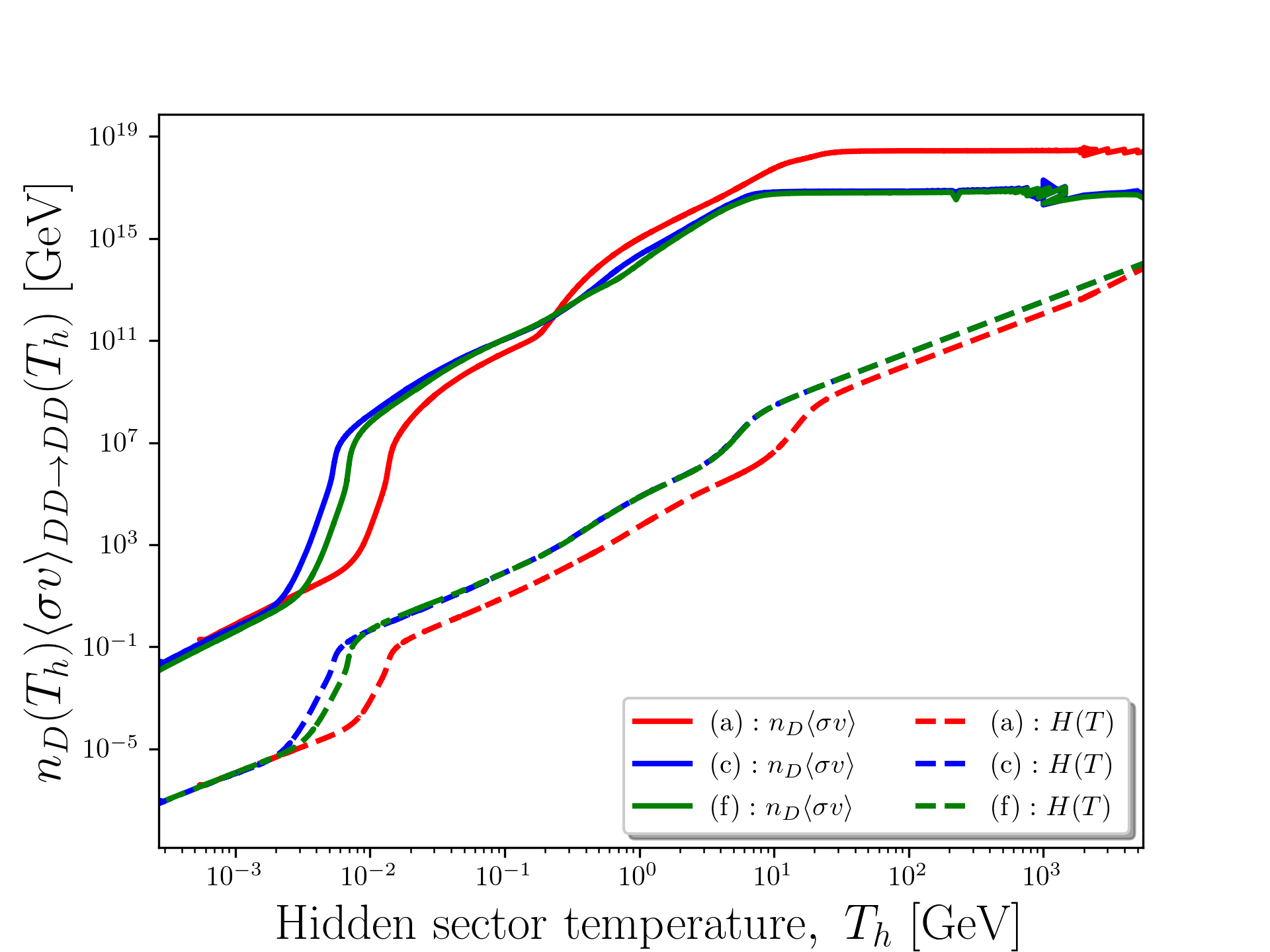} 
\caption{A plot of the dark matter self-interaction cross-section and the Hubble parameter for three benchmarks of Table~\ref{tab1}. One notices that $n_D\langle\sigma v\rangle$ remains higher than $H(T)$.}
\label{fig2}
\end{figure}

 In Fig.~\ref{fig1} we exhibit the dark freeze-out where the decoupling between the dark photon and the dark fermion, i.e. $n_{D}(T_h)\langle\sigma v\rangle_{D\bar{D}\to\gamma'\gamma'}(T_h)\sim H(T)$ occurs for values of $T_h/m_D\sim\mathcal{O}$(1$-$7)$\times 10^{-3}$ exhibited by the knee in the lower part of the plot.
The dark fermions interact with each other via the exchange of a dark photon or a $Z$ boson. The coupling of $D$ with the former is proportional to $g_X$ while its coupling with the latter is proportional to the gauge kinetic mixing $\delta$. Since $g_X$ is quite sizable, the $D$ fermions immediately enter into self-equilibrium after production and remain so even at low temperatures. This is shown in  Fig.~\ref{fig2} which is a plot of $n_D\langle\sigma v\rangle_{DD\to DD}$ (solid curves) and the Hubble parameter $H(T)$ (dashed curves) versus the hidden sector temperature. One can clearly see that the self-interaction processes ($DD\to DD$, $D\bar D\to D\bar D$ and $\bar D\bar D\to\bar D\bar D$) remain above $H(T)$ and thus in equilibrium even at low temperatures. This justifies the use of thermal averaging of cross-sections in the dark sector.  

\begin{figure}
 \includegraphics[width=0.49\textwidth]{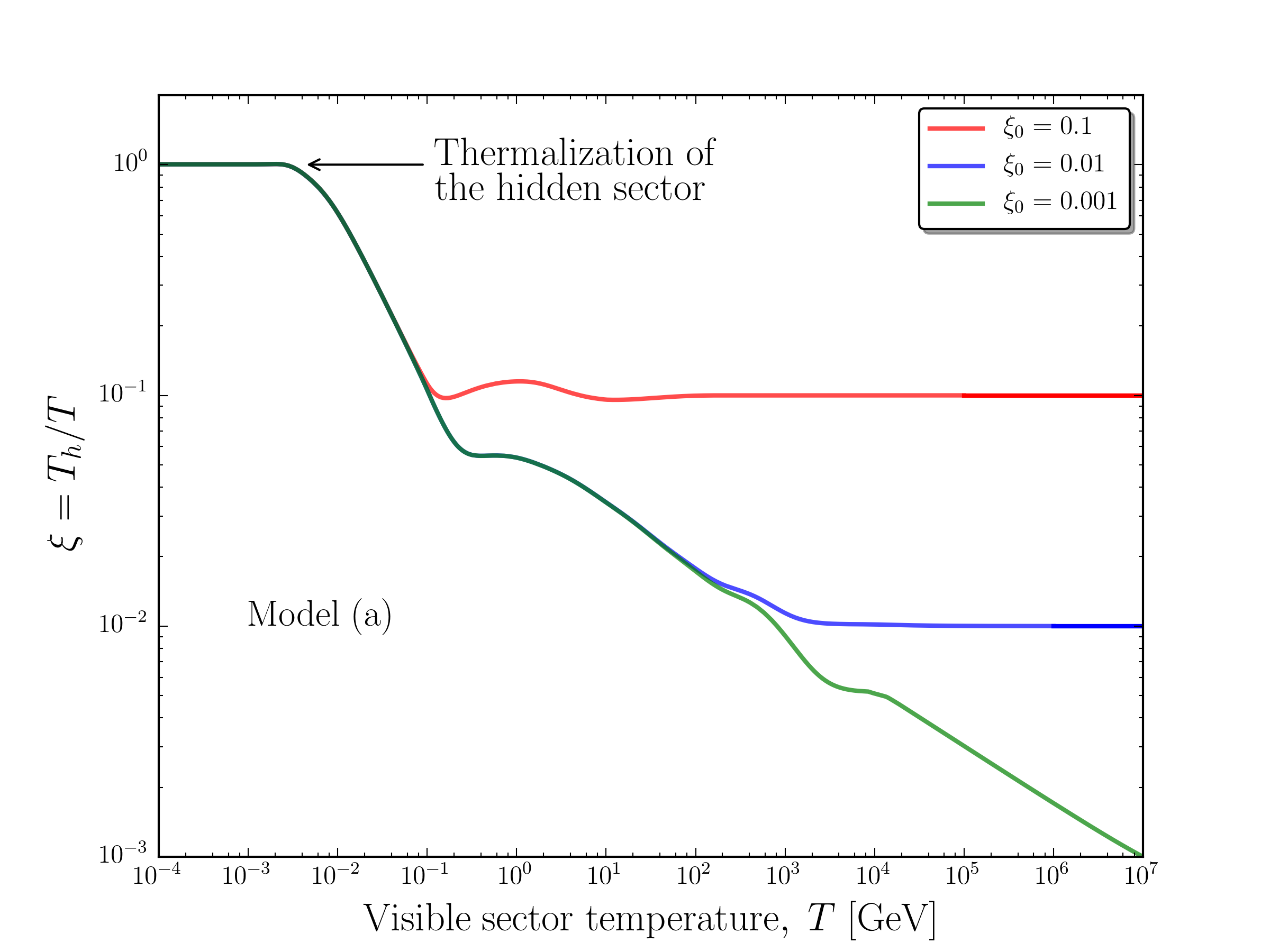} 
 \caption{Evolution of $\xi$ as a function of $T$ for  benchmark (a) of Table~\ref{tab1}
  for three different initial values of $\xi$ at high temperature.}
\label{fig3}
\end{figure}

\begin{figure}
 \includegraphics[width=0.45\textwidth]{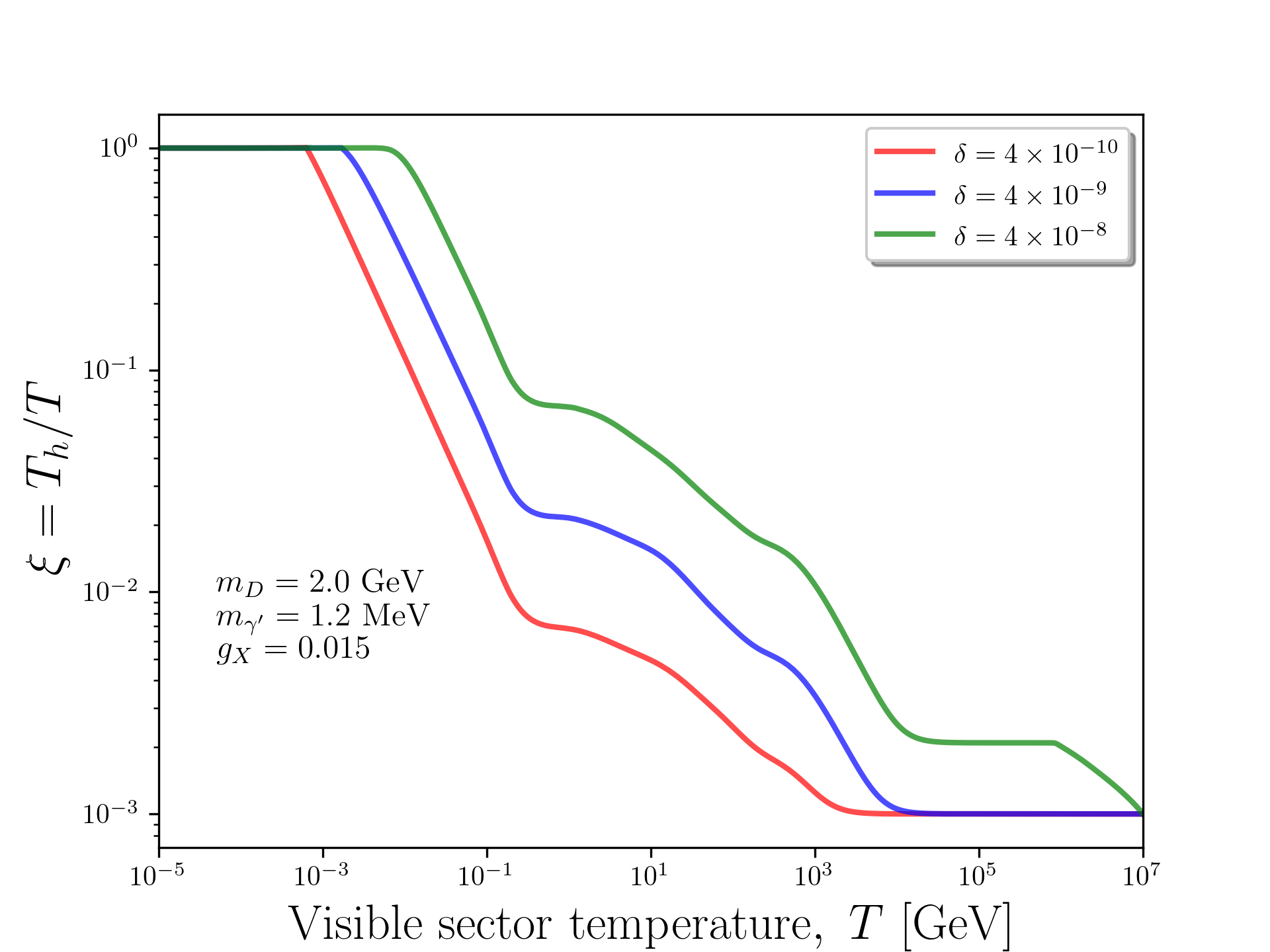} 
 \includegraphics[width=0.45\textwidth]{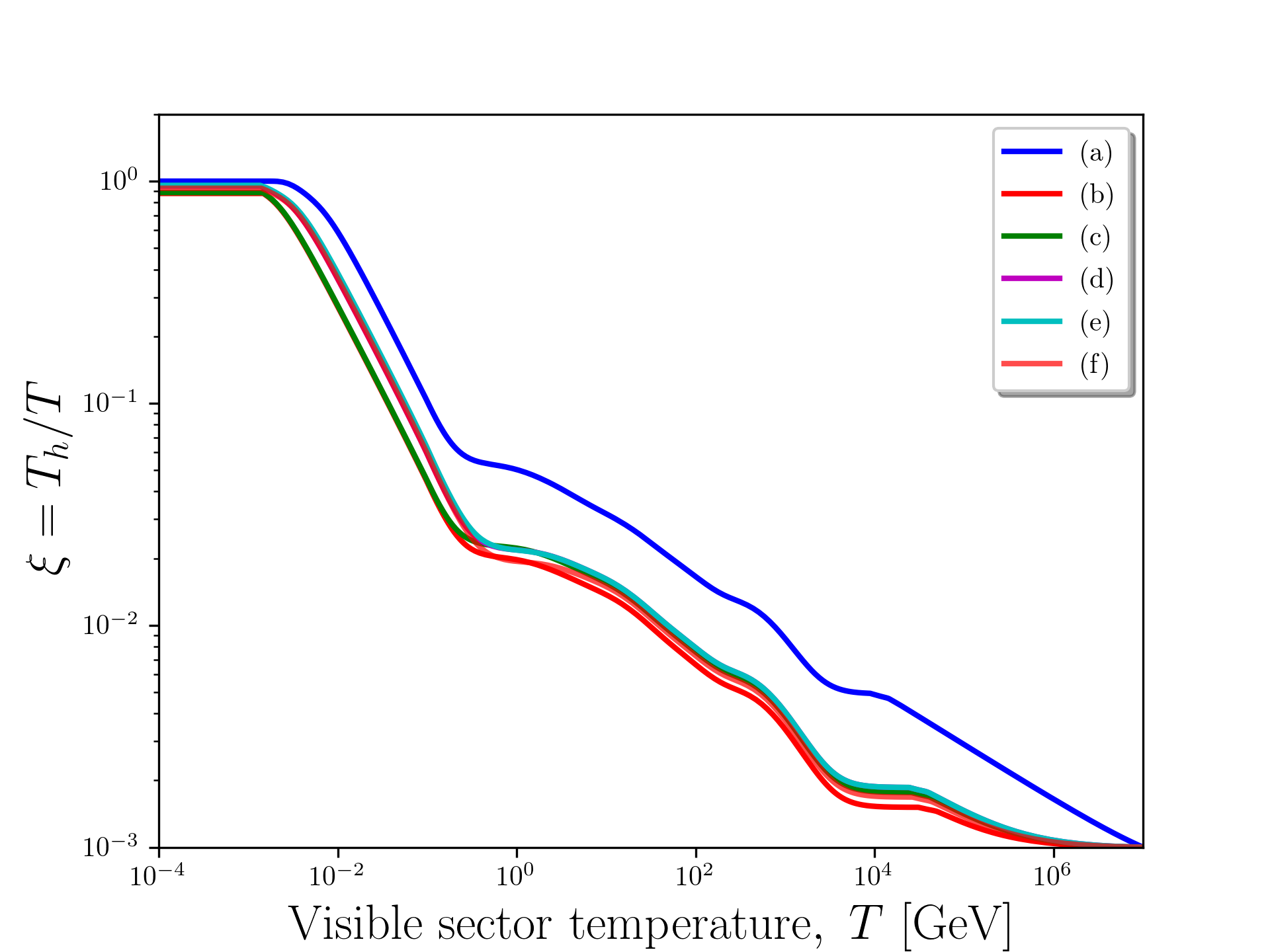} 
\caption{Evolution of $\xi$ with the visible sector temperature for three values of the gauge kinetic mixing $\delta$ (upper panel) and for the six benchmarks of Table~\ref{tab1} (lower panel). 
The upper panel exhibits the
phenomenon that the thermalization of the hidden and the visible sector takes  place at a lower temperature
 for smaller values of $\delta$ indicating that a more feeble coupling between the two sectors delays the thermalization process in contrast with a stronger coupling. 
 The lower panel shows that thermalization takes place for all model points of Table 1.}
\label{fig4}
\end{figure}

 In  Fig.~\ref{fig3}  we exhibit the phenomenon of thermalization of the hidden sector for one model point.
 Here one finds that starting with different  initial conditions on $\xi\equiv\eta^{-1}$ at some high temperature, 
 one ends up with  $\xi=1$, i.e., $T_h=T$ at low temperatures.
   We further discuss the thermalization of the hidden and visible sectors exhibited in Fig.~\ref{fig3}. 
First we note that we can look at the visible and hidden sectors as two heat baths. If there is 
a coupling between these two, they would eventually thermalize according to the second law
of thermodynamics. The rate at which they thermalize would be model dependent.
Thus thermalization could occur more rapidly (more slowly)  if the coupling between them is 
 stronger (weaker).  We exhibit this phenomenon in a quantitative fashion in Fig.~\ref{fig4}.
 Here we show that 
  thermalization happens for all the cases considered 
  but the time at which it happens depends on the coupling between the sectors which is parametrized by the kinetic mixing. To show this, we vary the kinetic mixing and plot the evolution of $\xi$ in the upper panel of Fig.~\ref{fig4}. We note  that for the three values of $\delta$, thermalization between the sectors eventually takes place but for larger couplings, thermalization sets in at higher temperature (green curve), i.e. earlier in time while for smaller couplings, thermalization takes place at a later stage, i.e. at lower temperatures (red curve). The same plot is given for the six benchmarks in the lower panel which shows the same observation. 

\begin{figure}
 \includegraphics[width=0.49\textwidth]{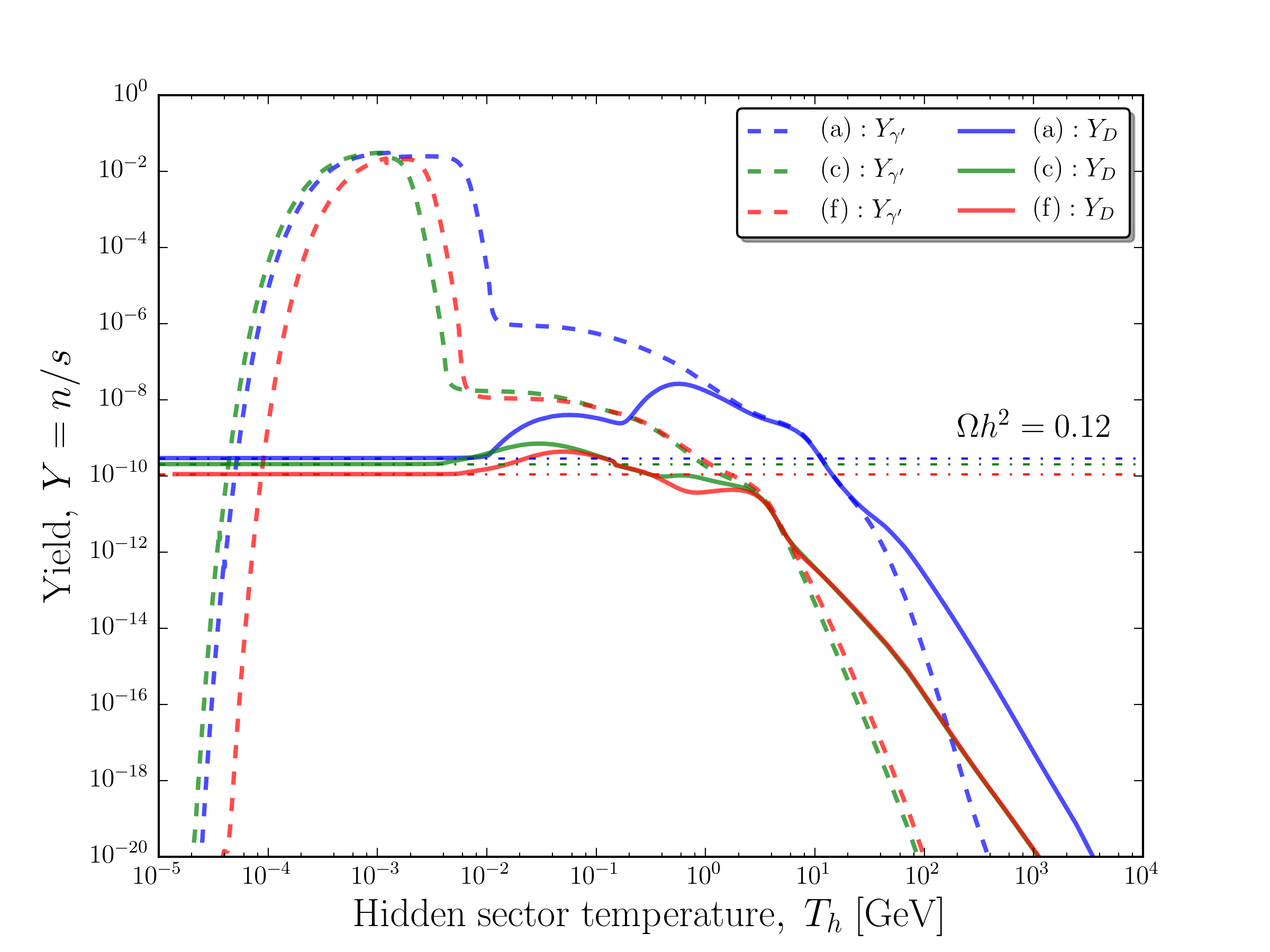}
    \caption{
  Evolution of $Y_D$ and $Y_{\gamma'}$  as a function of  $T_h$ for three benchmarks of Table~\ref{tab1}. The dashed horizontal lines correspond to the yields which give a relic density$~\sim 0.12$ consistent with Planck experiment~\cite{Aghanim:2018eyx} for each dark matter mass.}
\label{fig5}
\end{figure}

\begin{figure}
   \includegraphics[width=0.49\textwidth]{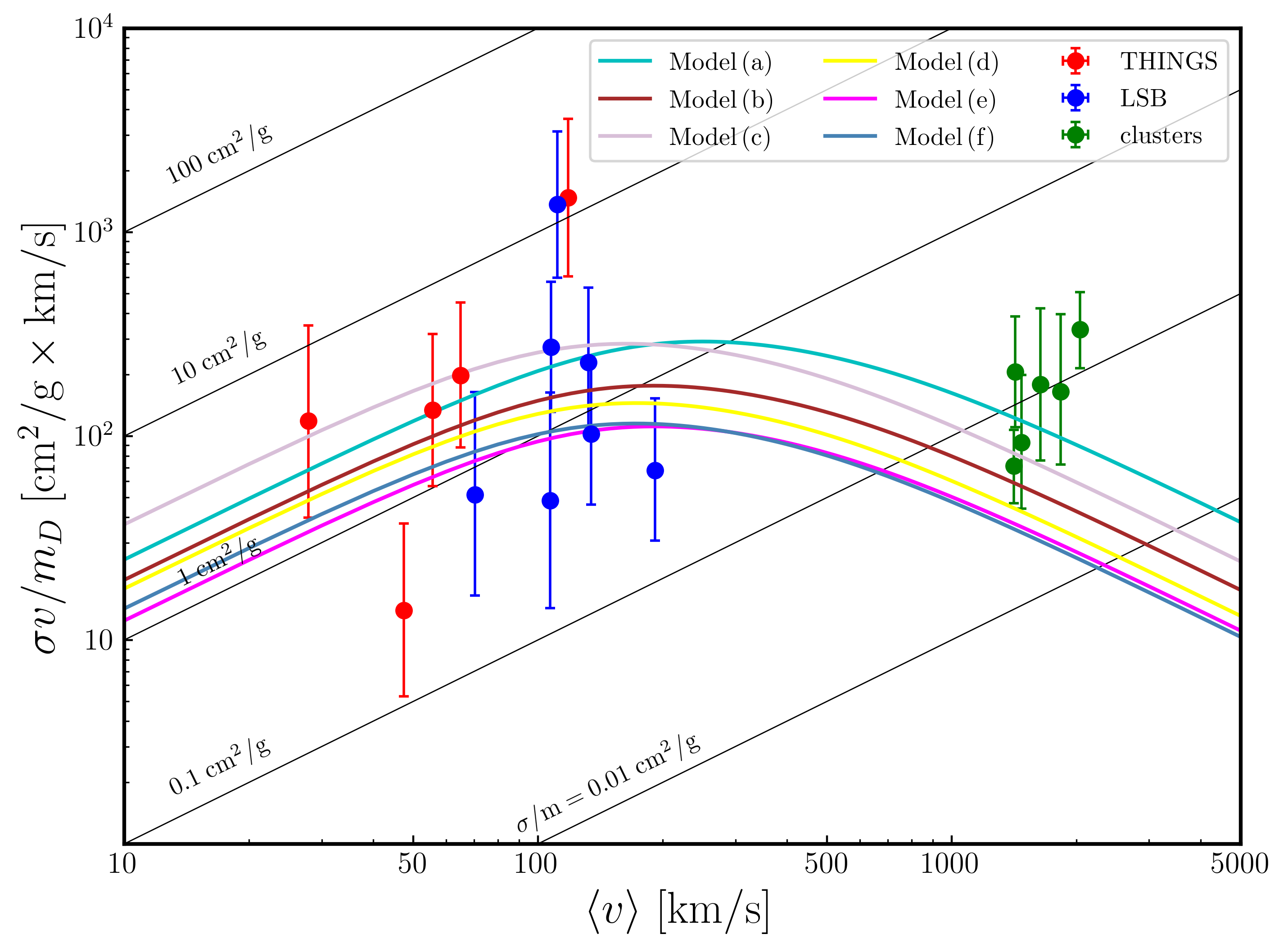}
   \caption{$\sigma v/m_D$  plotted versus $\langle v\rangle$ in the halo
    using self-interacting  dark matter cross-section for the six model points of Table~\ref{tab1}.
   The data points are taken from the work of ~\cite{Kaplinghat:2015aga,Sagunski:2020spe}.}     
	\label{fig6}
\end{figure}

The evolution of the yield for the dark fermions and dark photon in terms of the hidden sector temperature is shown in  Fig.~\ref{fig5} for three benchmarks of Table~\ref{tab1}. The injection of particle number density into the hidden sector from the visible sector is evident from the steep rise of the yield of $D$ (solid curve) and $\gamma'$ (dashed curve) showing the freeze-in mechanism at play. Once the hidden sector is populated enough, the processes $D\bar D\longleftrightarrow\gamma'\gamma'$ become important. This can be seen in Fig.~\ref{fig1} where the solid curves rise above the Hubble parameter $H(T)$ (dashed line) at high temperature. As the temperature drops, the process $D\bar D\to\gamma'\gamma'$ falls below $H(T)$ and the $2\to 2$ processes producing the dark fermions become less efficient. This causes the dark fermion number density to freeze-out as shown in  Fig.~\ref{fig5}. The increase in $\gamma'$ number density is sustained by the $2\to 1$ processes until the process $\gamma'\to e^+e^-$ dominates causing a dramatic drop in the dark photon number density. Thus the dark photons do not contribute to the relic density as they decay before the BBN. This shows that the mechanism behind producing the correct relic density is a combination of freeze-in due to the feeble couplings between the hidden and visible sectors and a dark freeze-out owing to the size of $g_X$ which give weak scale interactions in the dark sector. Though it can be minimal, the evolution of $\xi$ or $\eta$ has an effect on the relic density. For the benchmarks of Table~\ref{tab1}, we notice a change in the relic density by a factor of $\sim$ 2$-$3 when switching between $\xi_0=1$ (sectors have the same temperature) and $\xi_0=1000$ (starting with a cooler hidden sector).    
In  Fig.~\ref{fig6}  we give a plot of $\sigma v/m_D$ where $\sigma$ refers to self-interaction 
 cross-section and $v$ is the Moller velocity. The theory curves are for six model points of Table~\ref{tab1} using THINGS and LSB galaxies and clusters' analysis taken from~\cite{Kaplinghat:2015aga,Sagunski:2020spe}, showing that the models can fit the dark matter 
 cross sections from galaxy scales to clusters.
 
\begin{figure}
 \includegraphics[width=0.5\textwidth]{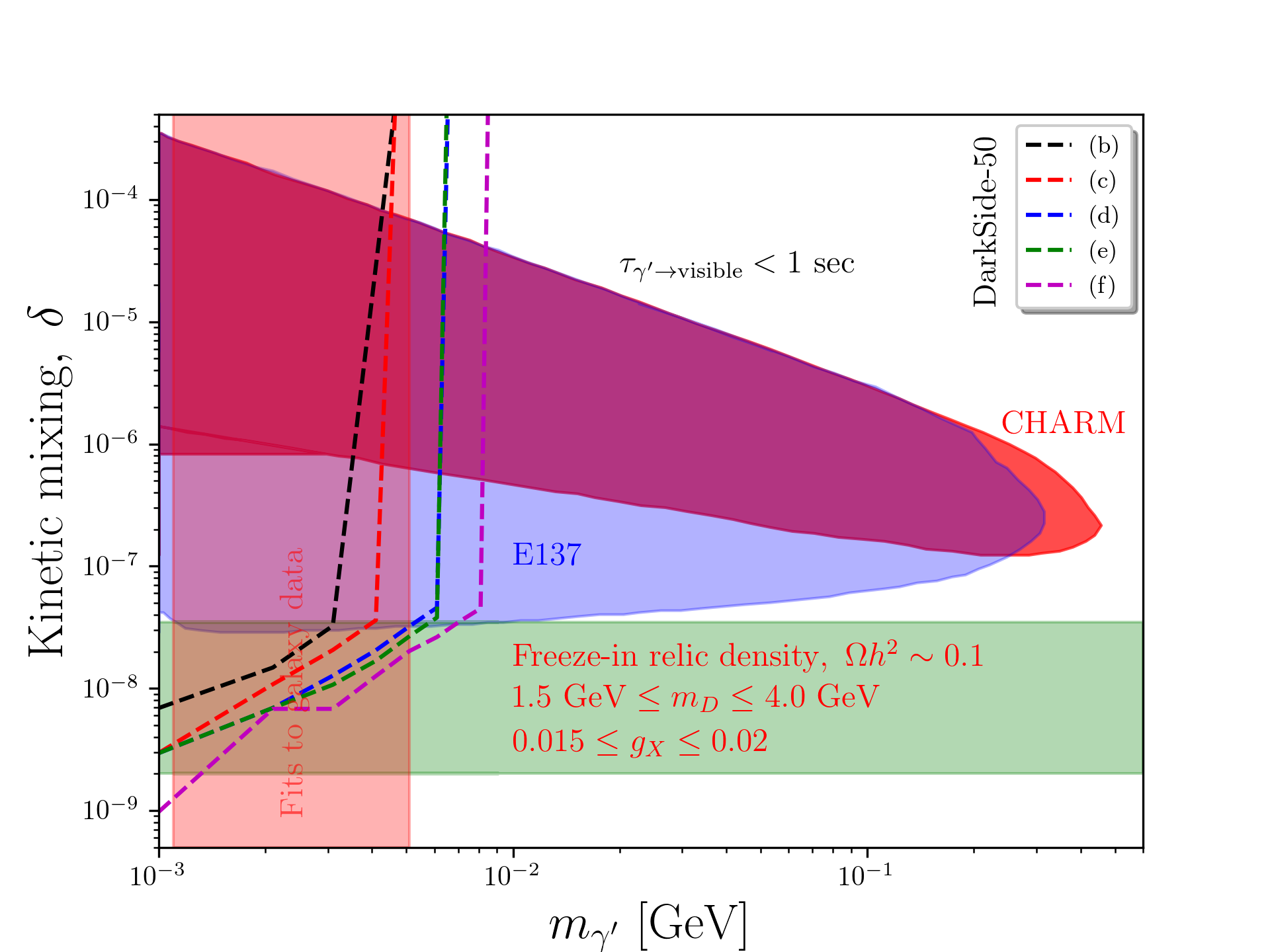} 
\caption{Plots in the gauge kinetic mixing-dark photon mass plane showing the allowed regions of the parameter space for a range of dark fermion mass and coupling $g_X$. The blue and red regions represent limits on dark photon decay to visible SM particles from experiments E137~\cite{Bjorken:1988as} and CHARM~\cite{Gninenko:2012eq,Bergsma:1985is}. The limit on the SI cross-section from DarkSide-50~\cite{Agnes:2018ves} is recast to our model and is shown as a dashed lines. The green band is the region giving the correct relic density from freeze-in and the red vertical band shows the region giving a fit to the DGC data.}
\label{fig7}
\end{figure}

\begin{widetext}

 \begin{figure}
   \includegraphics[width=0.32\textwidth]{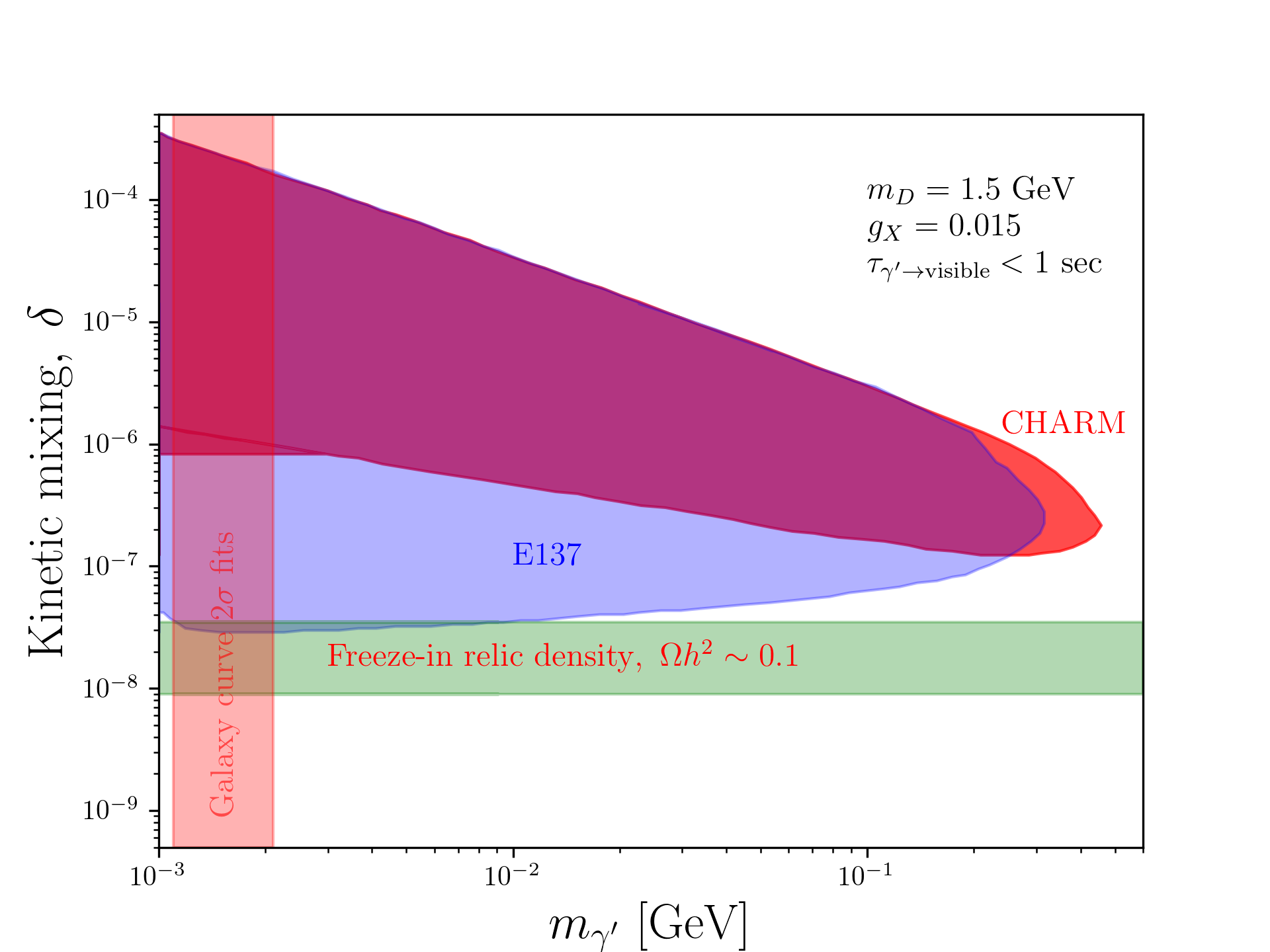} 
   \includegraphics[width=0.32\textwidth]{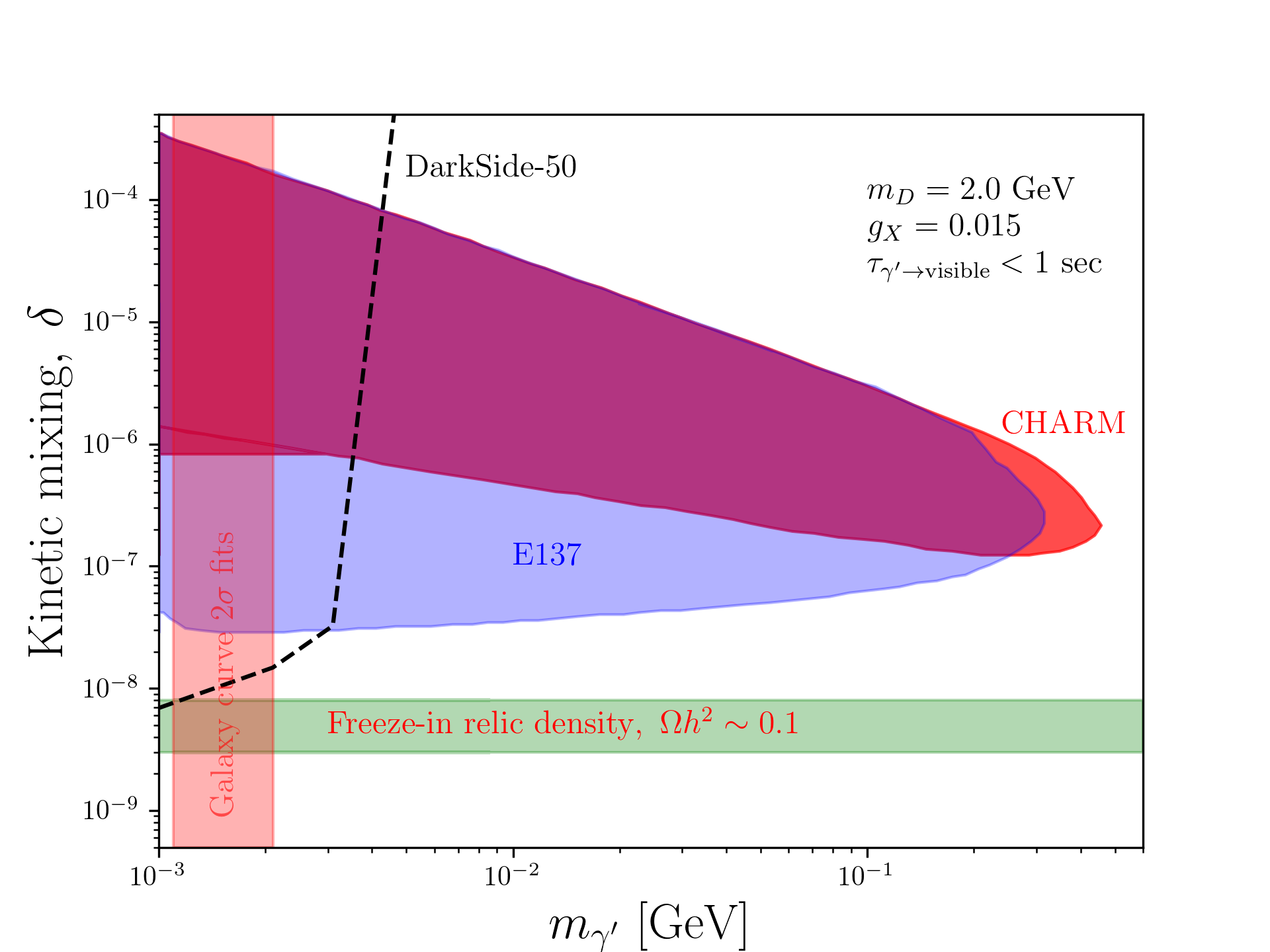}
  \includegraphics[width=0.32\textwidth]{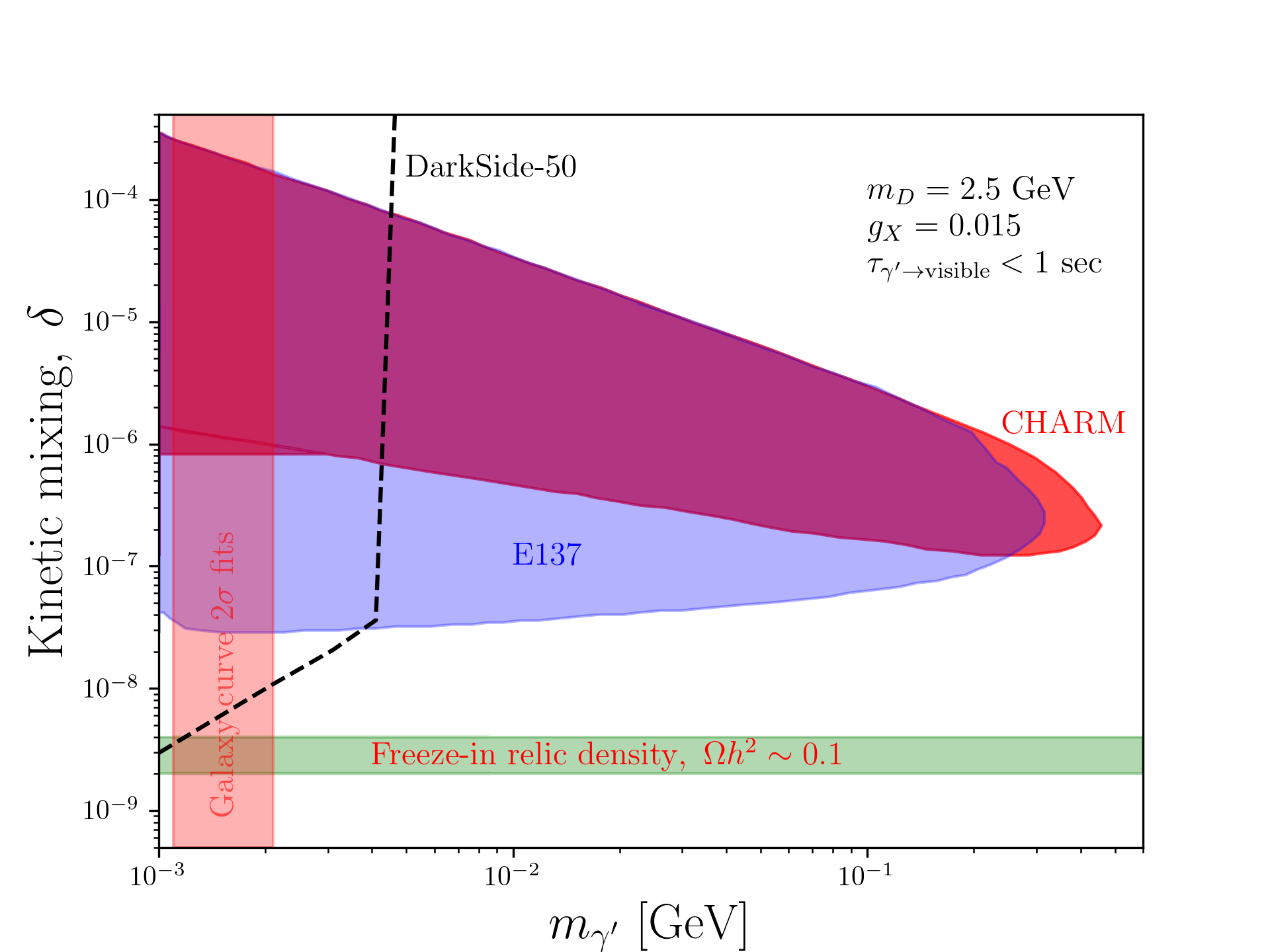} \\  
  \includegraphics[width=0.32\textwidth]{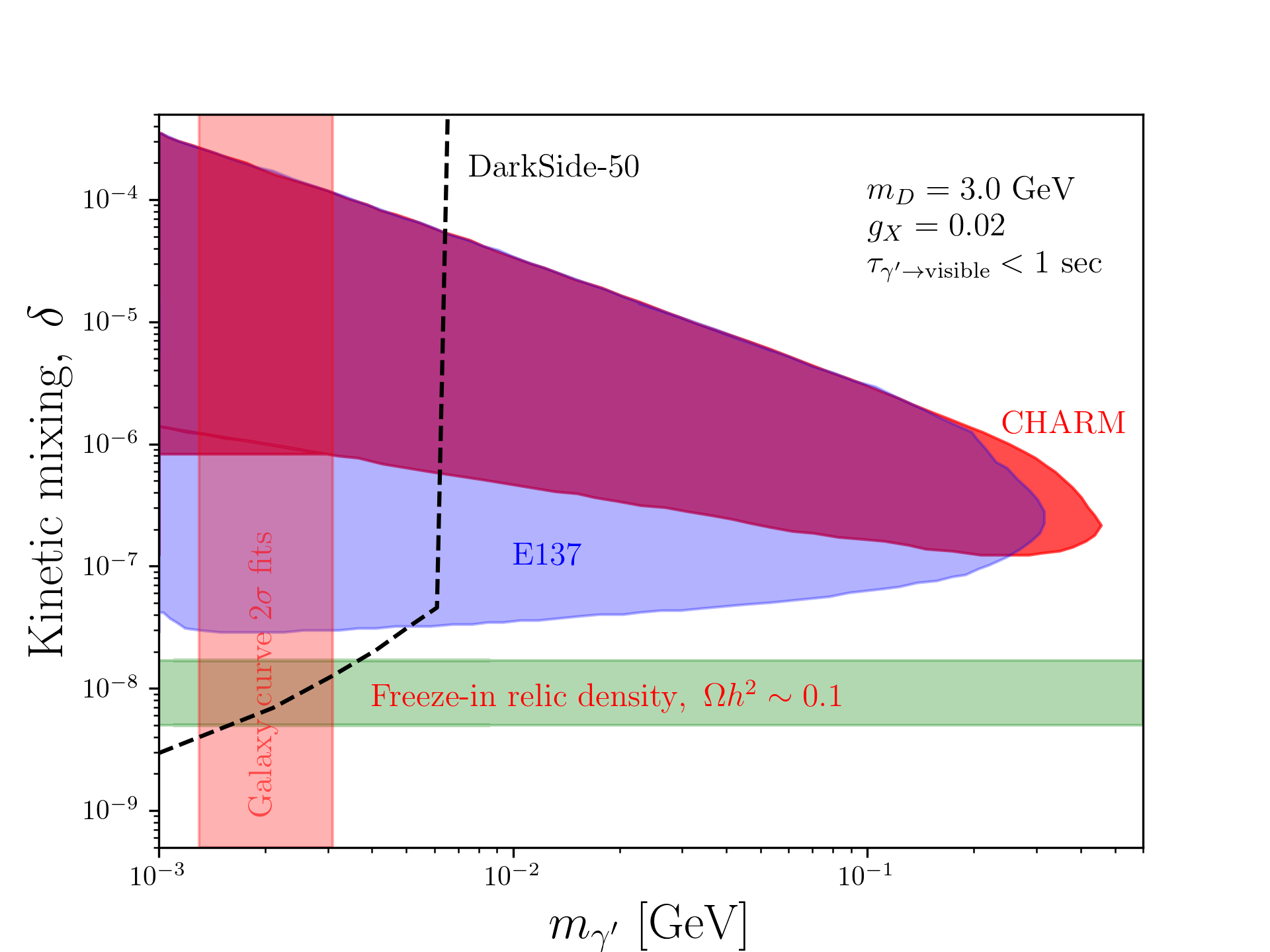} 
   \includegraphics[width=0.32\textwidth]{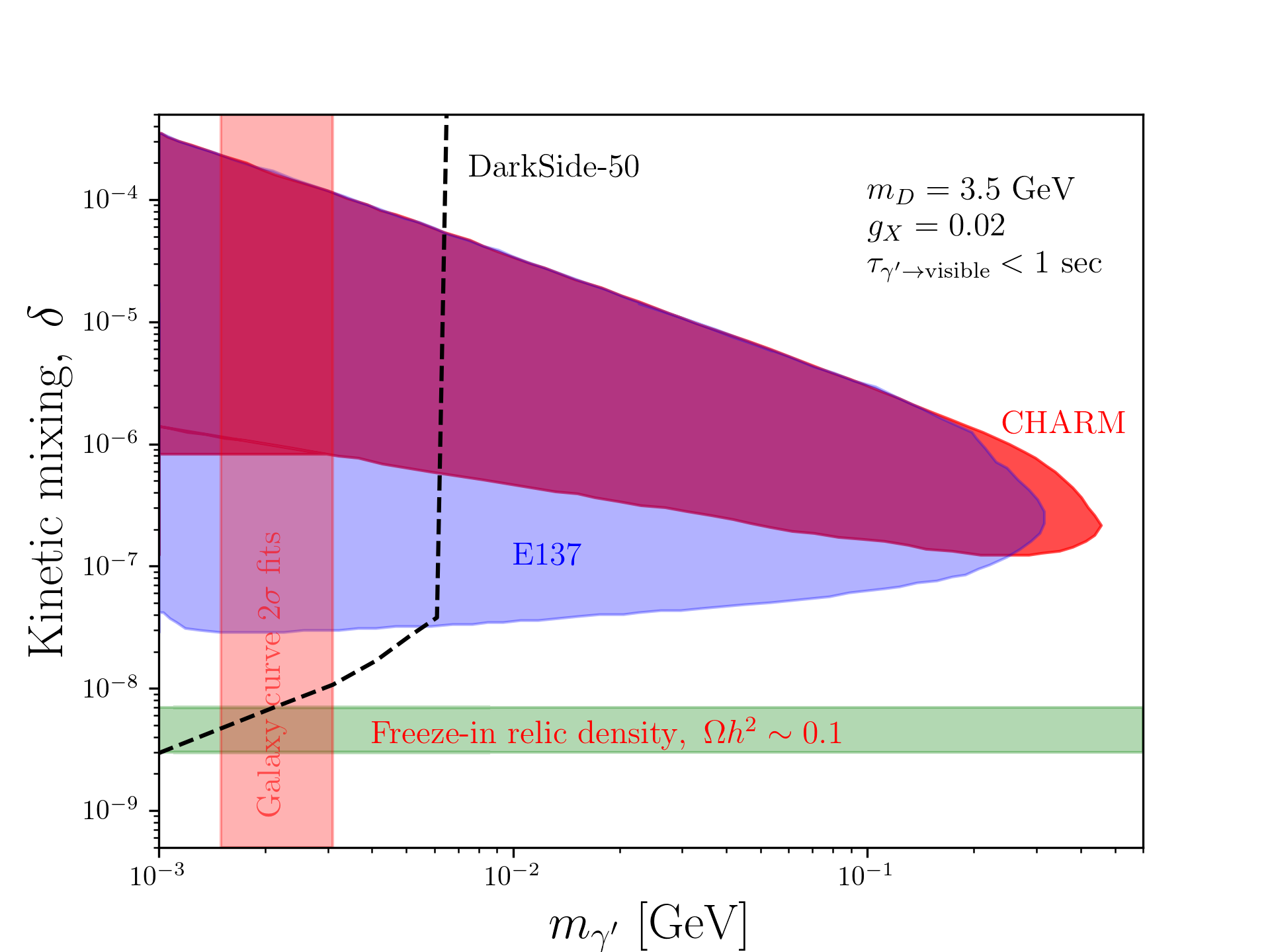}
  \includegraphics[width=0.32\textwidth]{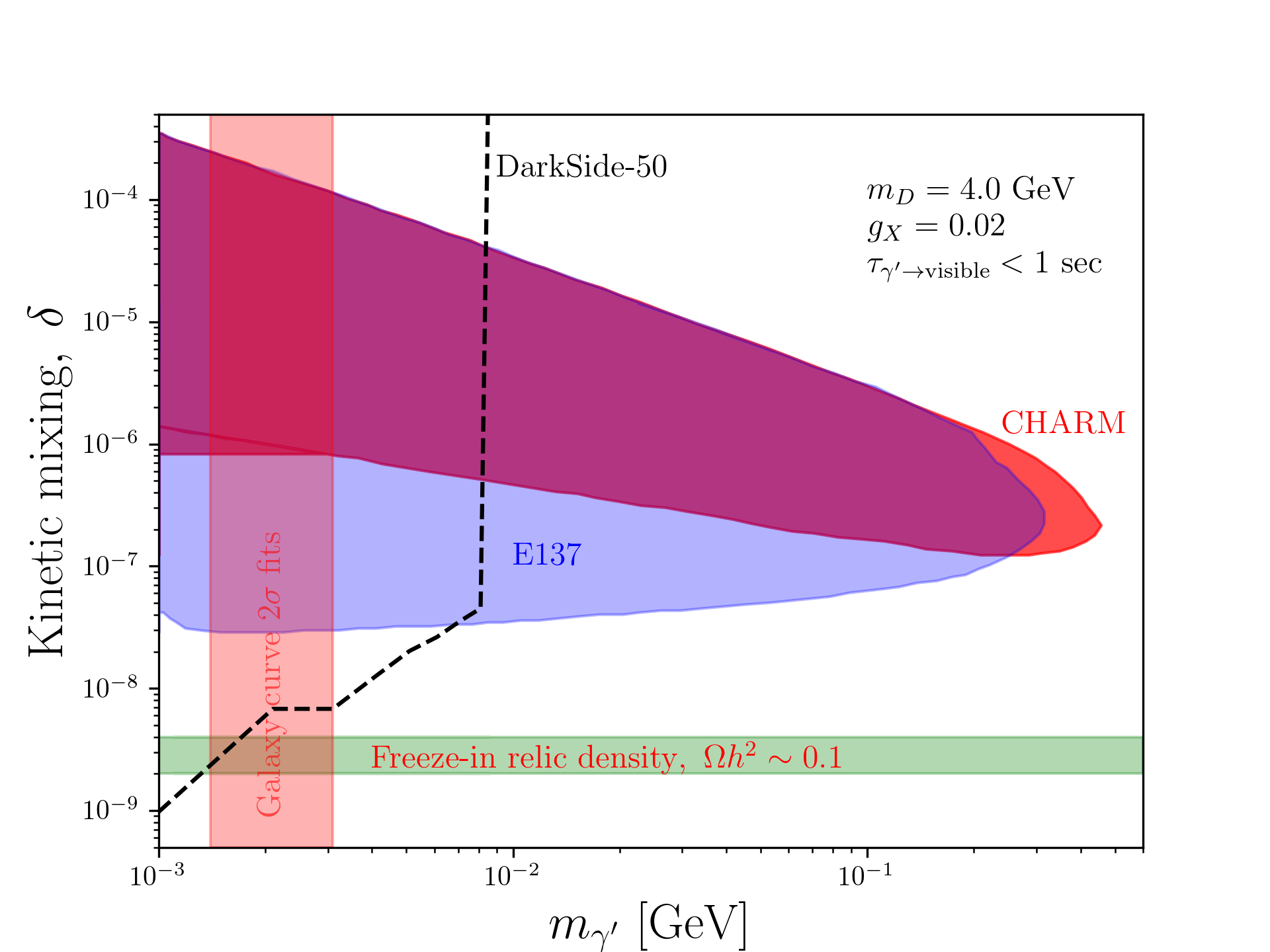} 
    \caption{Plots in the $\delta$-$m_{\gamma'}$ plane showing the allowed regions of the parameter space for six values of the dark fermion mass, $m_D$ and coupling $g_X$. The limits are the same as the ones described in the caption of Fig.~\ref{fig7}.}
\label{fig8}
\end{figure} 

\end{widetext}

The six benchmarks presented in Table~\ref{tab1} are only part of a larger parameter space where one can satisfy the dark matter relic density and produce a fit to the cross-sections from DGC data. To illustrate this,
we consider six values of the dark fermion mass $m_D=1.5,2.0,2.5,3.0,3.5$ and 4.0 GeV and two values of the couplings $g_X=0.015$ and 0.02 and for each set of $(m_D,g_X)$ we vary the gauge kinetic mixing 
$\delta$ and the dark photon mass $m_{\gamma'}$ and plot the allowed regions in the parameter space. The combined plot which includes the considered dark fermion mass range is given in Fig.~\ref{fig7}. The plot shows regions which satisfy the relic density constraints and give a fit to the DGC data, and are consistent with other laboratory and astrophysical constraints. 
We now discuss these constraints. These include constraints from dark photon experiments which are numerous~\cite{Essig:2013lka} and we only show the relevant and most stringent ones for our case, namely, from E137~\cite{Bjorken:1988as} (blue region) and CHARM~\cite{Gninenko:2012eq,Bergsma:1985is} (red region) which look at the decay of dark photons into visible Standard Model particles. Constraints on spin-independent proton-dark matter scattering cross-section from DarkSide-50 is recast to fit our model and is shown as dashed lines for each benchmark of Table~\ref{tab1}. There is no constraint from DarkSide-50 on the $m_D=1.5$ GeV case which can also be seen from Fig.~\ref{fig7}. The green horizontal band shown in Fig.~\ref{fig7} represents the region which produces the correct relic density from freeze-in and the vertical red band is the allowed region in which a good fit to the DGC data can be produced, within a $2\sigma$ corridor. We notice that there is a 
an allowed region where the relic density and galaxy fits are satisfied while escaping constraints from DarkSide-50 and dark photon experiments. This corresponds to a dark photon mass in the range $\sim$ 1$-$5 MeV, a gauge kinetic mixing $\delta\sim\mathcal{O}(10^{-9}$--$10^{-8})$ and $0.015\leq g_X \leq 0.02$ for $1.5~\text{GeV} \leq m_D\leq 4.0~\text{GeV}$. A plot similar to Fig.~\ref{fig7} is made for six sets of $(m_D, g_X)$ values and shown in Fig.~\ref{fig8}.

\begin{figure}
      \includegraphics[width=0.49\textwidth]{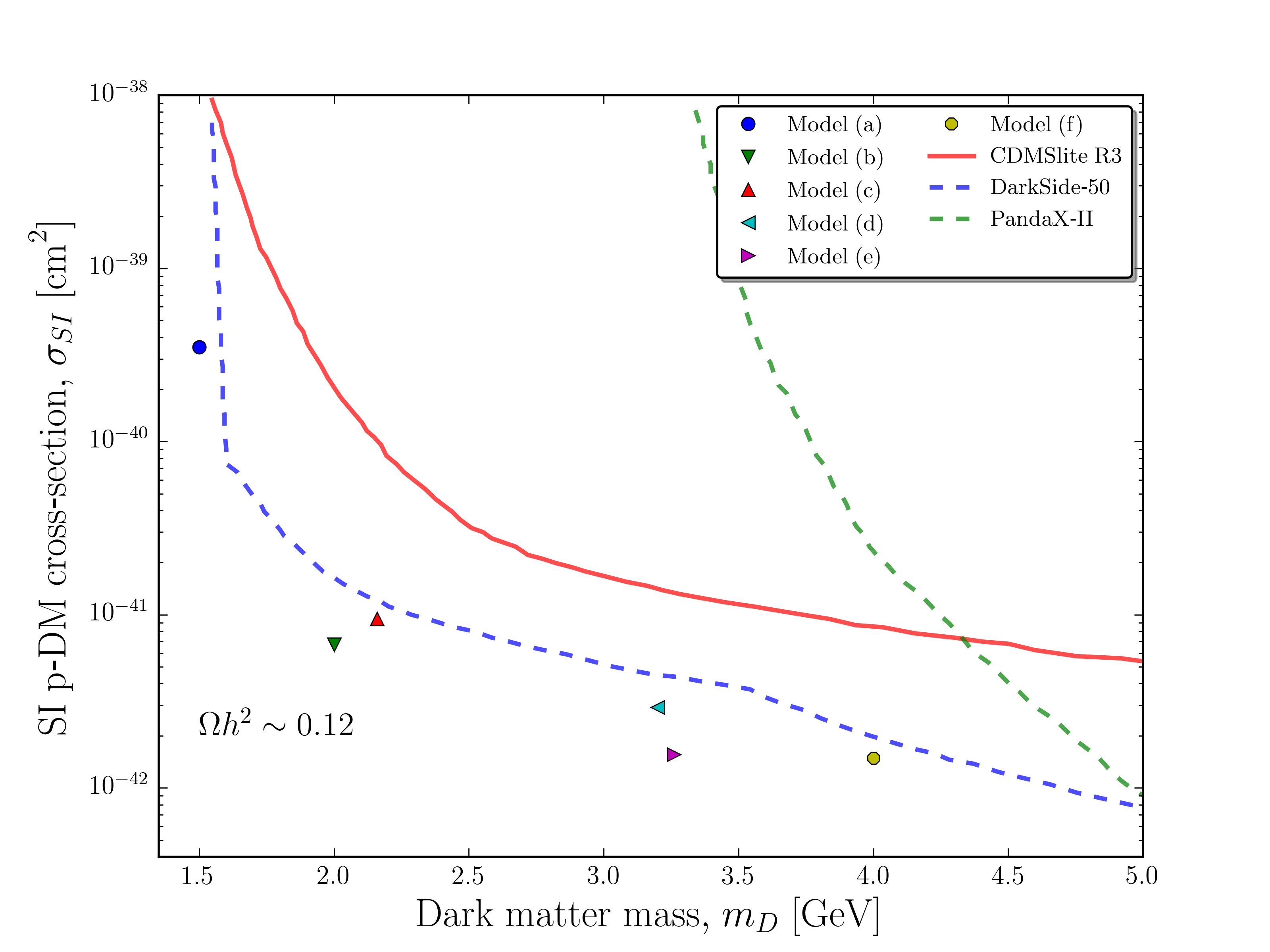}
   \caption{The spin-independent proton-DM scattering cross-sections for the six benchmarks of Table~\ref{tab1} calculated using \code{micrOMEGAs 5.0}~\cite{Belanger:2018ccd} with model files generated by \code{SARAH}~\cite{Staub:2013tta,Staub:2015kfa}. 
    Also shown are the current exclusion limits from CDMSlite R3,  DarkSide-50 and PandaX-II.}
	\label{fig9}
\end{figure}

  Finally, in Fig.~\ref{fig9} we exhibit the spin-independent 
 p-DM cross-section as a function of the dark matter mass $m_D$ where the current limits from CDMSlite R3~\cite{Agnese:2018gze}, DarkSide-50~\cite{Agnes:2018ves} and  PandaX-II~\cite{Tan:2016zwf} are also exhibited. One finds that the model points are consistent with the current limits including constraints from CMB~\cite{Bernal:2019uqr} and dark photon experiments~\cite{Essig:2013lka,Bjorken:1988as,Gninenko:2012eq,Bergsma:1985is} and can be explored in future improved experiments.
We note that while the model discussed above can resolve the cusp-core and too big to fail anomalies,
 a resolution of the missing satellites anomaly requires a very late kinetic decoupling. Using the formalism
 of~\cite{Bringmann:2009vf} we estimate  the kinetic decoupling temperature  to be $\mathcal{O}(100)$ keV. A further reduction to achieve very late kinetic decoupling 
      could be accomplished by the inclusion of more dark degrees of freedom 
      as discussed in~\cite{Bringmann:2018lay}   
   and the references therein.

\section{Conclusion}
New analytic results of this work are the three coupled equations  defined by Eqs.~(\ref{y1})$-$(\ref{y7}) which allow one to solve the Boltzmann equations 
 for the relic density of dark matter where the evolution depends on two temperatures, one for the hidden and the other for 
 the visible sector. It is then seen that one must simultaneously  evolve the ratio $\eta=T/T_h$ consistently
 to solve for the relic density. The analysis shows that thermalization of the hidden sector occurs for all the model points and the more feeble the interaction is the  longer it takes for thermalization
  to occur.
The hidden sector model we consider consists of a dark fermion $D$ and a dark 
  photon $\gamma'$ as mediator where the dark photon is unstable and decays before BBN. We present a set of model 
  points which satisfy the relic density constraint and their self-interactions produce velocity dependence of dark
  matter cross-sections within SIDM framework using DGC data.
   We note that the velocity dependence of dark matter cross sections is a direct consequence 
  of a force mediator mass in the range $\mathcal{O}$(MeV) and the confirmation of such velocity dependence
  would point to the existence of a dark force.
  The model points can be tested in future direct detection experiments
 via the spin-independent  p-DM cross-sections. We note that a confirmation of the velocity dependence of the DM cross section within the self 
 interacting
 dark matter model would point to the existence of a dark force mediated by a light dark photon 
 which controls the dynamics of dark matter from galaxy scales to scales of galaxy clusters.

We discuss now the analysis of this work in the context of previous works.
Thus as noted above one of the main results of this work are the set of equations, Eqs.~(\ref{y1})$-$(\ref{y7}).  It is widely realized in the literature (see, e.g.,~\cite{Chu:2011be} or~\cite{Hambye:2019dwd}) 
that a  proper treatment of coupled visible and hidden sectors which are not
in thermal equilibrium requires the evolution of the ratio of the visible and hidden sector temperatures. However, an explicit  set of equations that accomplish this  does not exist in the literature. The work of~\cite{Chu:2011be}  gives a broad analysis of four ways of creating dark matter.
  In this work the dark photon is assumed massless. However, a massless dark photon cannot produce a Yukawa-like force that is needed to produce a velocity dependence of dark matter cross sections which
  we discuss in this work.  Further,  while this work recognizes the importance of evolution
  of the ratio of the visible and hidden sector temperatures, no explicit equation for the evolution of the
  ratio of two temperatures, i.e., the analogue of Eq.~(\ref{y3}), is given.
  In~\cite{Hambye:2019dwd}, the dark photon is given a mass and the paper  discusses the importance of 
  a proper treatment of  two temperatures $T$ and $T'$ in the evolution. 
  However, the closest work comes to how $T'$ is to be determined is Eq.~(3.25) which is not
  an explicit differential equation such as Eq.~(\ref{y3}) of our work. Further, we note that in~\cite{Hambye:2019dwd}   as well 
  as in~\cite{Chu:2011be}, aside from the absence of explicit analytic formula on $T'$ vs $T$, there is also  
   no numerical exhibition of the evolution of the ratio $\xi=T'/T$ while this is done in Fig.~\ref{fig3} and Fig.~\ref{fig4}. We note that the analysis of~\cite{Hambye:2019dwd} deals with millicharges which is also not directly relevant to our work. 
  In our analysis the  hidden sector  equilibrates with itself. This is shown to manifest in that a freeze-out
  is achieved in this sector as exhibited in Fig.~\ref{fig1} of the paper. We note that the deviations from equilibrium must be accounted for as discussed in~\cite{Binder:2017rgn}
  for the case of  freeze-out and in~\cite{DEramo:2020gpr} for the case of freeze-in.
 \\

We thank Sean Tulin for a communication. 
The analysis presented here was done using the resources of the Momentum Cluster at Northeastern University. WZF was supported in part by the National Natural Science Foundation of China under Grant No. 11905158 and No. 11935009. The research of AA, PN and ZYW was supported in part by the NSF Grant PHY-191332.

\section{Appendix}
The $J$-functions that appear in Eq.~(\ref{y6}) are defined as 
\begin{align}
&n^{\rm eq}_i(T)^2 J(i~\bar{i}\to D\bar{D})(T)\nonumber\\
&=\frac{T}{32\pi^4}\int_{s_0}^{\infty}ds~\sigma_{D\bar{D}\to i\bar{i}}s(s-s_0)K_2(\sqrt{s}/T), \\
&n^{\rm eq}_i(T)^2 J(i~\bar{i}\to \gamma')(T)\nonumber\\
&=\frac{T}{32\pi^4}\int_{s_0}^{\infty}ds~\sigma_{i\bar{i}\to \gamma'}s(s-s_0)K_2(\sqrt{s}/T), 
\end{align}
\begin{equation}
n_{\gamma'}J(\gamma'\to e^+ e^-)(T_h)=n_{\gamma'}m_{\gamma'}\Gamma_{\gamma'\to e^+ e^-},
\end{equation}
and
\begin{align}
&n_i^{\rm eq}(T)^2\langle\sigma v\rangle_{i\bar{i}\to\gamma'}(T) \nonumber\\
&= \frac{T}{32\pi^4}\int_{s_0}^{\infty} ds ~\sigma(s) \sqrt{s}\, (s-s_0)K_1(\sqrt{s}/T),
\end{align}
where $K_1$ is the modified Bessel function of the second kind and degree one and $s_0$ is the minimum of the Mandelstam variable $s$.
The self-interaction cross-sections for $D\bar D\to D\bar D$,  $DD\to DD$, and $\bar D \bar D\to \bar D \bar D$ 
are given by 
\begin{equation}
\frac{d\sigma}{d\Omega}=\sum_{i=1}^3 \frac{\overline{|\mathcal{M}_i|^2}}{64\pi^2s},
\end{equation}
where for $D\bar D\to D\bar D$ 
\begin{align}
\overline{|\mathcal{M}_1|^2}&=2g_X^4\Bigg\{\frac{t^2+u^2+8m_D^2s-8m_D^4}{(s-m_{\gamma'}^2)^2+\Gamma^2_{\gamma'} m_{\gamma'}^2}\nonumber \\
&+\frac{u^2+s^2+8m_D^2t-8m_D^4}{(t-m_{\gamma'}^2)^2}\nonumber \\
&+\frac{2[m_{\gamma'}^4-m_{\gamma'}^2(s+t)+st+\Gamma^2_{\gamma'}m^2_{\gamma'}]}{[m_{\gamma'}^4-m_{\gamma'}^2(s+t)+st]^2} \nonumber \\
&\times (u^2-8m_D^2u+12m_D^4)\Bigg\}.
\end{align}
For $DD\to DD$  
\begin{align}
\overline{|\mathcal{M}_2|^2}&= 2 g_X^4\Bigg\{\frac{s^2+u^2-8m_D^2(s+u)+24m_D^4}{(t-m_{\gamma'}^2)^2}\nonumber\\
&+\frac{t^2+s^2-8m_D^2(s+t)+24m_D^4}{(u-m_{\gamma'}^2)^2}\nonumber\\
&+\frac{2[m_{\gamma'}^4-m_{\gamma'}^2(u+t)+ut+\Gamma^2_{\gamma'}m^2_{\gamma'}]}{[m_{\gamma'}^4-m_{\gamma'}^2(u+t)+ut]^2}\nonumber\\
&\times(s^2-8m_D^2s+12m_D^4)\Bigg\},
\end{align}
where $s,t,u$ are the Mandelstam variables.
For $\bar D\bar D\to \bar D \bar D$,
   $\overline{|\mathcal{M}_3|^2} = \overline{|\mathcal{M}_2|^2}$.  
The cross-section for the process $D\bar D\to \gamma' \gamma'$ is given by 
\begin{align}
&\sigma^{D\bar{D}\to {\gamma'} {\gamma'}}(s)=\frac{g_{X}^4(\mathcal{R}_{11}-s_{\delta}\mathcal{R}_{21})^4}{8\pi s(s-4m^2_D)}\nonumber\\
&\times\Bigg\{-\frac{\sqrt{(s-4m^2_{{\gamma'}})(s-4m^2_{D})}}{m^4_{{\gamma'}}+m^2_D(s-4m^2_{{\gamma'}})}[2m^4_{{\gamma'}}+m^2_D(s+4m^2_D)] \nonumber \\
&+\frac{\log A}{s-2m^2_{{\gamma'}}}(s^2+4m^2_D s+4m^4_{{\gamma'}}-8m_D^4-8m_D^2 m^2_{{\gamma'}})\Bigg\},
\label{DDgg}
\end{align}
with
\begin{equation}
A=\frac{s-2m^2_{{\gamma'}}+\sqrt{(s-4m^2_{{\gamma'}})(s-4m^2_{D})}}{s-2m^2_{{\gamma'}}-\sqrt{(s-4m^2_{{\gamma'}})(s-4m^2_{D})}}.
\end{equation}
Here $\mathcal{R}_{11}$ and $\mathcal{R}_{21}$ are matrix elements of $\mathcal{R}$ which diagonalizes the mass and kinetic 
energy matrices as given in~\cite{Feldman:2007wj}.
When kinematically allowed the process  $\gamma'\gamma'\to D\bar D$ is given by 
\begin{equation} 
9(s-4m_{\gamma'}^2)\sigma^{\gamma'\gamma'\to D\bar D}(s) = 8(s-4m_D^2)\sigma^{D\bar{D}\to {\gamma'} {\gamma'}}(s).
\end{equation}

\end{document}